\documentclass[12pt,a4paper]{article}

\usepackage[latin1]{inputenc}
\usepackage{amsmath}
\usepackage{graphicx}
\usepackage{verbatim}

\newcommand{\bc}{\begin{center}}
\newcommand{\ec}{\end{center}}
\newcommand{\be}{\begin{equation}}
\newcommand{\ee}{\end{equation}}
\newcommand{\bea}{\begin{eqnarray}}
\newcommand{\eea}{\end{eqnarray}}
\newcommand{\beqn}{\begin{eqnarray}}
\newcommand{\eeqn}{\end{eqnarray}}
\newcommand{\ba}{\begin{array}}
\newcommand{\ea}{\end{array}}
\newcommand{\bal}{\begin{aligned}}
\newcommand{\eal}{\end{aligned}}
\newcommand{\ben}{\begin{enumerate}}
\newcommand{\een}{\end{enumerate}}
\newcommand{\bitem}{\begin{itemize}}
\newcommand{\eitem}{\end{itemize}}

\newcommand{\fr}{\frac}

\newcommand{\crn}{\nonumber \\}

\newcommand{\noi}{\noindent}

\newcommand{\eq}[1]{Eq.~(\ref{#1})}
\newcommand{\bib}[1]{\cite{#1}}
\newcommand{\bibs}[1]{\cite{#1}}
\newcommand{\fig}[1]{Fig.~\ref{#1}}
\newcommand{\tab}[1]{Table~\ref{#1}}
\newcommand{\sect}[1]{Section~\ref{#1}}
\newcommand{\ssect}[1]{Subsection~\ref{#1}}

\newcommand{\gev}{{\unskip\,\text{GeV}}}
\newcommand{\tev}{{\unskip\,\text{TeV}}}
\newcommand{\mev}{{\unskip\,\text{MeV}}}

\newcommand{\fb}{{\unskip\,\text{fb}}}

\newcommand{\lt}{{\texttt{LoopTools}}}
\newcommand{\fc}{{\texttt{FormCalc-6.0}}}
\newcommand{\fa}{{\texttt{FeynArts-3.4}}}
\newcommand{\sloops}{{\texttt{SloopS}}}

\newcommand{\doc}{{\texttt{D0C}}}
\newcommand{\OL}{{\texttt{OneLOop}}}

\newcommand{\bases}{{\texttt{BASES}}}

\newcommand{\vegas}{{\texttt{VEGAS}}}

\newcommand{\qd}{{\texttt{QD}}}

\newcommand{\epem}{e^+e^-}

\newcommand{\eezzz}{e^+e^-\to ZZZ}
\newcommand{\eezzzt}{$e^+e^-\to ZZZ\;$}
\newcommand{\eezzztp}{$e^+e^-\to ZZZ$}
\newcommand{\eezzzgam}{e^+e^-\to ZZZ\gamma}
\newcommand{\eewwz}{e^+e^-\to W^+W^- Z}
\newcommand{\eewwzt}{$e^+e^-\to W^+W^- Z \;$}
\newcommand{\eewwztp}{$e^+e^-\to W^+W^- Z $}
\newcommand{\eewwzgam}{e^+e^-\to W^+W^- Z\gamma}
\newcommand{\eewwt}{$e^+e^-\to W^+W^-  \;$}

\newcommand{\hs}{\hspace*{3mm}}

\newcommand{\eg}{{\it e.g.}}

\newcommand{\dsreal}{\ensuremath{\text{d}\sigma_{\text{real}}}}
\newcommand{\dssub}{\ensuremath{\text{d}\sigma_{\text{sub}}}}
\newcommand{\dsborn}{\ensuremath{\text{d}\sigma_{\text{Born}}}}

\def\slashepi{\epsilon_i\kern -.720em {/}}
\def\slashpi{p_i\kern -.600em {/}}

\begin{document}

\begin{titlepage}

\vspace*{0.1cm}\rightline{MPP-2009-214}
 \rightline{LAPTH-1372/09}

\vspace{1mm}
\begin{center}

{\Large{\bf  NLO corrections to \eewwzt and \eezzzt }}

\vspace{.5cm}

Fawzi Boudjema$^a$, Le Duc Ninh$^b$, Sun Hao$^a$ and Marcus M. Weber$^b$

\vspace{4mm}
$^a${\it LAPTH, Universit\'e de Savoie, CNRS,  \\
BP 110, F-74941 Annecy-le-Vieux Cedex, France}

$^b${\it Max-Planck-Institut f\"ur Physik
(Werner-Heisenberg-Institut), \\ D-80805 M\"unchen, Germany}

\vspace{10mm} \abstract{We calculate the one-loop electroweak
corrections to \eewwzt and \eezzzt and analyse their impacts on
both the total cross section and some key distributions. These
processes are important for the measurements of the quartic
couplings of the massive gauge bosons which can be a window on the
mechanism of spontaneous symmetry breaking. We find that even
after subtracting the leading QED corrections, the electroweak
corrections can still be large especially as the energy increases.
We compare and implement different methods of dealing with
potential instabilities in the routines pertaining to the loop
integrals. For the real corrections we apply  a dipole subtraction
formalism and compare it to a phase-space slicing method.}

\end{center}

\normalsize

\end{titlepage}

\section{Introduction}

The LHC has just started running again and seems now to be on
course for what it has been built for: discovery of the last
remaining particle of the much successful Standard Model, the
Higgs boson. It may well be that before this particle is uncovered we
will have seen clear signs of New Physics that better
encompasses an elementary Higgs boson. The conventional Minimal
Supersymmetric Standard Model is the most popular example of such
a scenario. It may however happen that  the mechanism of
electroweak symmetry breaking will remain elusive and that one has
to look for subtle deviations in standard processes. Because of
its clean environment a linear collider might be more suited
for this purpose.

 From this perspective the
study of \eewwzt and \eezzzt may be very instructive and would
play a role similar to $\epem \to W^+W^-$ at lower energies.
Indeed it has been stressed  that  \eewwzt and \eezzzt  are prime
processes for probing the quartic vector boson couplings
\cite{Belanger:1992qh}. In particular deviations from the gauge
value in the quartic $W^+W^-ZZ$ and $ZZZZ$\footnote{$ZZZZ$ is
absent at tree-level in the Standard Model. Other photonic quartic
couplings may be probed in the processes we study but they are
best studied in other reactions. Moreover the latter, from the
point of view symmetry breaking, are less relevant and would be of
higher order \cite{Belanger:1999aw}.} couplings that are accessible
in these reactions might be the residual effect of physics
intimately related to electroweak symmetry breaking. Since these
effects can be small and subtle, knowing these cross sections with
high precision is mandatory. This calls for theoretical
predictions taking into account loop corrections.

Apart from the physics
motivations for performing such calculations, the other reason is
that one-loop corrections, in particular the electroweak
corrections, for such $2 \to 3$ processes are a good testing
ground for the various ingredients and techniques that enter such
one-loop multi-leg corrections. Although recently NLO corrections to $2 \to 4$
processes have set the technical frontier with a handful of
processes in this category having been addressed\footnote{As far
as electroweak corrections are concerned only two such
calculations have been
performed \cite{Boudjema:2005rk,Denner:2005fg}.}, NLO corrections
to $2\to 3$ processes are far from straightforward.
Not only the number of diagrams differs greatly from
one process to another but perhaps more importantly the loop
structure can also differ significantly. For the processes at hand
one has to deal with high rank tensors, rank-$4$, for the pentagon
diagrams, compared to at most a rank-2 for $\epem \to\nu \bar \nu
H$ \cite{Belanger:2002me,Belanger:2002ik,Denner:2003iy,Denner:2003yg}.
This might lead to much more severe numerical instabilities due to the
appearance of higher powers of the inverse Gram determinants in
the tensor reduction. Moreover different scales and masses may lead to
sensitive issues related to Landau singularities in scalar integrals
 \cite{ninh_bbH2}. It
is therefore important to conduct one-loop corrections to a variety
of $2 \to 3$ processes.

Radiative corrections to \eezzzt have appeared recently in
\cite{JiJuan:2008nn} and those to \eewwzt in \cite{Wei:2009hq}
while we were preparing this paper.  We have made an independent
calculation of the electroweak corrections to \eewwzt and \eezzzt.
Preliminary results on \eewwzt have been presented in
\cite{ninh-corfu} before those of \cite{Wei:2009hq} were made
public. We perform two independent calculations and check further
through non-linear gauge parameter independence tests. These help
also identify potential instabilities in the routines pertaining
to the loop integrals. We detail how some critical issues related
to inverse Gram determinants have been tackled, how real
corrections have been implemented and how checks were conducted.
Our calculation is implemented in two independent Monte Carlo
codes which can calculate total cross sections and arbitrary
distributions.  We compare our results with those in
\bibs{JiJuan:2008nn,Wei:2009hq} and comment also on the renormalisation
scheme and input parameters.

\section{Calculational details}
\label{sect-code}

\begin{figure}[t]
\begin{center}
\includegraphics[width=0.7\textwidth]{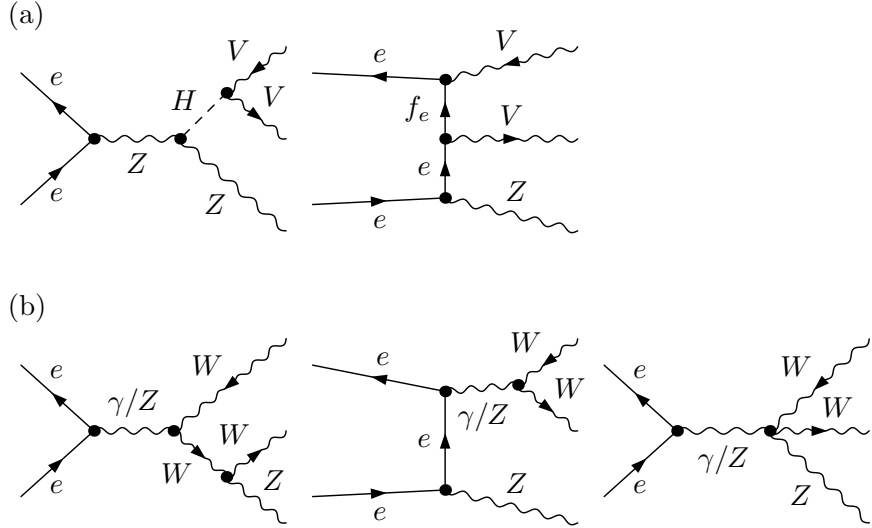}
\end{center}
  \caption{{\em Representative Born diagrams for $\eezzz$ and $\eewwz$. Diagrams (a) contribute to
    both processes while  diagrams of type (b) contribute only to $\eewwz$. The first diagram of type (a)
    will be referred to as the Higgsstrahlung contribution.}}
  \label{fig:treediag}
\end{figure}

At leading order $W^+W^-Z$ and $ZZZ$ final states are produced
through the diagrams shown in \fig{fig:treediag}. A common
contribution to the two processes is the Abelian-like t-channel
fermionic exchange akin to the QED process $\epem \to 3\gamma$.
Both processes include the Higgsstrahlung contribution where the
splitting $H^\star \to VV$ occurs. Apart from this contribution
\eewwzt can be built up from \eewwt through the addition of $Z$
radiation from either the initial or final state and the
$s$-channel quartic $WWZZ$. Since the precision electroweak data
suggest a Higgs mass below the $WW$ threshold, we restrict our study to the
region $M_H < 160\gev$.
This means that the Higgsstrahlung
contribution can not be resonant and therefore in our
calculation no width is introduced.

We also set the electron mass to zero whenever possible. The
electron mass then appears only in mass singular logarithms. These
arise in the virtual corrections from loop diagrams containing
electrons and from photons radiated off electrons in the real
corrections.

We have performed our calculation in at least two independent ways
both for the virtual and the real corrections leading to two
independent numerical codes. A comparison of both codes has shown
full agreement at the level of the integrated  cross
sections as well as all the distributions that we have studied.

The phase-space integration is done by using the Monte Carlo
integrator \bases~\cite{bases, Yuasa:1996kx} in one code while the
other code employs \vegas~\cite{Lepage:1977sw}.



\subsection{Renormalisation}
\label{sec:calc-ren}
 We adopt the on-shell renormalisation scheme
as detailed in \cite{grace, Denner:1991kt}. By default, in
this scheme the electromagnetic coupling is defined in the Thomson
limit at $q^2 \to 0$. The counterterm $e \to Y e=(1+\delta Z_e) e$
is related through a Ward identity to the wave function
renormalisation constant of the photon and the wave function
describing the $A \to Z$ transition defined at $q^2=0$ so that the
photon propagator is defined with residue equal to one and no $A
\to Z$ transition remains when the photon is on shell~\cite{grace,
Denner:1991kt}. In the conventions of~\cite{grace} this leads to

\begin{eqnarray}
\delta Z_e= -\delta Z^{1/2}_{AA}+\frac{s_W}{c_W} \delta
Z^{1/2}_{ZA}, \quad   \delta Z^{1/2}_{AA}=\frac{1}{2} \frac{{\rm
d}}{{\rm d} q^2} \Pi_T^{AA}(q^2)_{|q^2=0}, \quad \delta
Z^{1/2}_{ZA}=-\frac{1}{M_Z^2} \Pi_T^{AZ}(0). \label{dze0}
\end{eqnarray}
$\Pi_T^{VV}$ is the transverse parts of the $VV$ self-energy. This
particular definition of the charge at the scale $q^2=0$  is not
the most appropriate since the weak processes take place at scales
of order $M_Z$ or higher. The running of $\alpha$ from $q^2=0$ to
$q^2=M_Z^2$ alone amounts to a $6\%$ correction. For a process of
order $\alpha^n$, this running will amount to a correction of
order $n\times 6\%$, thus hiding more interesting corrections.
Moreover these corrections due to the running are sensitive to the
light fermion masses through logarithms of the type $\ln
(q^2/m_f^2)$. In fact the effective couplings of the $Z$ to
fermions are also sensitive to isospin breaking effects and
therefore  virtual heavy top effects through $\Delta
r$~\cite{Sirlin:1980nh}. The combined effect of these two
corrections is better parameterised if one uses the Fermi coupling
{\it in lieu} of $\alpha(0)$. We will therefore use a variant of
the $G_\mu$ scheme. This scheme absorbs a large
universal part of $\mathcal{O}(\alpha)$ corrections into the Born
contribution. The advantage of the scheme is that the final
results are not sensitive to the light fermion masses, in
particular the light quark masses, and some universal $m_t^2$
corrections are also absorbed. In this scheme we
 use $\{G_\mu, M_Z, M_W, \text{other masses}\}$ instead
of $\{\alpha(0), M_Z, M_W, \text{other masses}\}$ as input
parameters, from which the electromagnetic coupling constant is
calculated as
\begin{eqnarray}
\alpha_{G_\mu}=\fr{\sqrt{2}G_\mu M_W^2}{\pi} s_W^2, \quad
s_W^2=\left(1-\fr{M_W^2}{M_Z^2}\right).
\end{eqnarray}

To avoid double counting we have to subtract the one-loop part of the
universal correction from the explicit ${\cal O}(\alpha)$ corrections
by using the counterterm

\begin{eqnarray}
\delta Z_e^{G_\mu}=\delta Z_e-\fr{1}{2}(\Delta r)_{1-{\rm  loop}}.
\end{eqnarray}
In the 't~Hooft-Feynman gauge with the usual linear gauge fixing
$(\Delta r)_{1-{\rm  loop}}$ is given by~\cite{Denner:1991kt, Sirlin:1980nh}
\begin{eqnarray}
\bal (\Delta r)_{1-{\rm loop}}=& 2 \delta Z_e - \fr{\delta
s_W^2}{s_W^2} -\biggl( \fr{\Pi_T^{WW}(0)+\delta M_W^2}{M_W^2}
\biggr) -
\fr{2}{s_W c_W} \delta Z^{1/2}_{ZA} \\
&+\fr{\alpha}{4\pi s_W^2}\left[6+\fr{7-4s_W^2}{2s_W^2}\ln
c_W^2\right]. \eal
\end{eqnarray}

Using $\alpha_{G_\mu}$ in tree-level calculations will therefore take
care of some universal higher-order contributions. In the $G_\mu$
scheme the corrections are initially of order
$\alpha_{G_\mu}^4$. Considering however that the typical scale for
both virtual photon exchange and real photon radiation is $q^2=0$ we
rescale our results so that the NLO results are of order
$\alpha_{G_\mu}^3 \alpha(0)$.



\subsection{Virtual corrections}

The virtual corrections have been evaluated using a conventional
Feynman-diagram based approach using standard techniques in the
two independent codes. The total number of diagrams in the 't~Hooft-Feynman
gauge is about 2700
including 109 pentagon diagrams for $\eewwz$ and about 1800
including 64 pentagons for $\eezzz$. This already shows that
$\eewwz$ with as many as 109 pentagons is more challenging than
$\eezzz$.\\

\underline{Code 1} \\
The first code uses \fa\ \cite{Hahn:2000kx} to generate all
Feynman diagrams and amplitude expressions. \fc\
\cite{Hahn:1998yk, Hahn:2009bf} is used to simplify and generate a
{\tt Fortran 77} code suited for the numerical evaluation of the
differential cross sections. We also use \sloops\
\cite{Boudjema:2006kn,Baro:2008bg,Baro:2009gn} an automated code
that uses a few modules from \fa\ but which implements the
generalised non-linear gauge (NLG) \cite{Boudjema:1995cb,grace}
\begin{eqnarray}
\label{fullnonlineargauge} {{\cal L}}_{GF}&=&-\frac{1}{\xi_W}
|(\partial_\mu\;-\;i e \tilde{\alpha} A_\mu\;-\;ig c_W
\tilde{\beta} Z_\mu) W^{\mu +} + \xi_W \frac{g}{2}(v
+\tilde{\delta} H +i \tilde{\kappa} \chi_3)\chi^{+}|^{2} \nonumber \\
& &\;-\frac{1}{2 \xi_Z} (\partial.Z + \xi_Z \frac{g}{ 2 c_W}
(v+\tilde\varepsilon H) \chi_3)^2 \;-\frac{1}{2 \xi_A} (\partial.A
)^2 \;.
\end{eqnarray}
The $\chi$ represents the Goldstone. We take the 't~Hooft-Feynman
gauge with $\xi_W=\xi_Z=\xi_A=1$ so that no ``longitudinal'' term
in the gauge propagators contributes. Not only does this make the
expressions much simpler and avoids unnecessary large
cancellations, but it also avoids the need for higher tensor
structures in the loop integrals. The use of the five parameters,
$\tilde{\alpha}, \tilde{\beta}, \tilde{\delta}, \tilde{\kappa},
\tilde\varepsilon $ is not redundant as often these parameters
check complementary sets of diagrams.  At a few random points in
phase space we exploit these parameters to perform powerful tests
on the generated matrix elements.  This test can reveal for
example numerical instabilities that are due to the reduction
algorithm in some points in phase space. While at a regular point
in phase space the
non-linear gauge check attains a 14 digit
agreement in double-precision when changing a NLG parameter from 0 to 1,
at non-regular points the same tests can fail.

Five-point one-loop integrals (up to rank 4) are reduced to
four-point integrals by using the reduction method of Denner and
Dittmaier \cite{Denner:2002ii}. By default, four-point and
three-point tensor integrals are reduced to scalar integrals by
using the Passarino-Veltman reduction
algorithm \cite{Passarino:1978jh}. With the latter we have observed   serious problems of
numerical instability related to four-point tensor integrals. This
occurs when the Gram determinants associated to these tensor
integrals, defined by $\det(G_3) = \det (2p_ip_j)$,
 become sufficiently small. We have
solved this problem in two  ways. In
the first method, we use a simple extrapolation trick using  the
segmentation technique described in \cite{Boudjema:2005hb} when
the Gram determinant is small enough. The condition implemented in
our code is
\begin{eqnarray}
\label{cdtgram} \fr{\det(G_3)}{(2p_{max}^2)^3}<10^{-7},
\end{eqnarray}
where $p_{max}^2$ is the maximum external mass of a box diagram.
In this limit the $N$-point function of rank $M$ is written as a
combination of $(N-1)$-point functions of rank $M$. This is done
directly at the level of the loop integral in momentum space
before introducing any Feynman parameters. This implementation
requires that one supplements the standard libraries with the
reduction of the tensors of rank $M=N+1$, for certain $N$-point
functions. Another way of tackling this problem in the first code
is to calculate all the loop integrals in quadruple precision. For
this study we have performed this everywhere in phase space and
not just for the points that satisfy \eq{cdtgram}. The
numerical integration becomes very stable even in the case of very
small Gram determinant. The price to pay is that the computation
speed is about 6 times slower than using the segmentation method.
We have obtained  agreement of cross sections within integration errors between the
segmentation method and using quadruple precision.

We have also observed that the scalar one-loop four-point integral
can show numerical problems and the library \lt\ \cite{looptools_5p, ff, ff0} alone is not good
enough for our calculations.
While reverting to quadruple precision (everywhere in \lt )
remedies the problem, in double precision we call other loop
libraries  to calculate scalar one-loop four-point integrals for
some special cases where \lt\ fails. \OL\ \cite{vanHameren:2009dr}
is used for some special cases with zero internal masses. Other
specific cases that we have identified are treated with
\doc\ \cite{Nhung:2009pm}, a code to calculate scalar one-loop four-point integrals
with complex/real masses. Generically, numerical instabilities in
scalar one-loop integrals can
originate from the following two sources. One is related to an
endpoint singularity manifested as a  pole very close to the
boundary of the integration interval. The other is called a pinch
singularity where there are two poles sitting very close to each
other with a very small imaginary part. These are both called
Landau singularities in Feynman loop integrals (\eg\ see
\cite{ninh_thesis, ninh_bbH2} for a detailed discussion). At
NLO, these singularities (up to four-point) are integrable but
they may cause numerical instability. For the case of the pinch
singularity the sign of the imaginary part of a
pole (in the complex plane)
can be wrongly calculated and hence a regular case where both
poles sitting on the same side of the real axis can be numerically
misidentified as a pinch singularity or vice versa. We have
observed that in our calculations, in particular the process
$\eewwz$, numerical problems in the scalar four-point integrals
are related to both sources. An efficient way to cross check the
results of scalar integrals is therefore to introduce a
tiny positive width for internal masses ($\Gamma_i=10^{-5}m_i$). In this
case, the masses become complex and the code \doc\ must be used.
This is indeed what we did in our calculations to obtain the
preliminary results \cite{ninh-corfu} which agree within
integration errors with our final results. In \eezzzt an example
where \doc\ is used is for $\sqrt{s}=300\gev$. Here the problem is
related to the $t\bar{t}$ threshold ($\sqrt{s}<2m_t$ in this case)
in the box diagram with four top quarks in the loop. Since all the
important discriminants in \doc\ depend only on
external momenta, the problem does not occur in \doc{}.

\underline{Code 2} \\
The second code also uses \fa\ for Feynman-diagram and
amplitude generation and \fc\ to evaluate the amplitudes. The
analytical output of \fc\ in terms of Weyl-spinor chains and
coefficients containing the tensor one-loop integrals is then
translated to C++ code after performing further optimizations of
the expressions.

The evaluation of the one-loop tensor integrals is done by reducing
them to a set of scalar integrals. The 5-point integrals are written
in terms of 4-point functions following \cite{Denner:2005nn},
which avoids leading inverse Gram determinants and the associated
numerical instabilities.  The remaining 3- and 4-point tensor
integrals are recursively reduced to scalar integrals with the
Passarino--Veltman algorithm \cite{Passarino:1978jh}.  For exceptional
phase-space points this reduction scheme becomes numerically
unstable. In this case we reevaluate both the scalar integrals and the
reduction itself in higher precision using the \qd\ library
\cite{Bailey:qd}. To determine when to switch to quadruple precision
we use the condition number\footnote{The condition number of a
  symmetric matrix is defined as the ratio of the largest and smallest
  eigenvalues, see {\it e.g.} \cite{Golub:1997}.} of the Gram matrix.
This is a good estimator of the number of digits lost
in the solution of the linear equation system appearing in
Passarino--Veltman reduction. While this simple estimator is
sufficient for triangle integrals in the case of 4- and 5-point
integrals the numerical instabilities can also originate from
small Gram determinants in the lower N-point integrals. We
therefore use not only the condition number of the N-point Gram
matrix but also the condition numbers of the (N-1)- and
(N-2)-point integrals appearing in the tensor reduction in these
cases.

Finally, the scalar integrals are calculated using the results of
\cite{hooft_velt, denner_d0} for the finite and
\cite{Beenakker:1988jr, Ellis:2007qk} for the IR singular
integrals. The scalar integrals and the tensor reduction have been
implemented as a C++ library allowing the calculation both in
double and higher precision. Internally the library uses
dimensional regularization for both UV and IR divergences. For
this calculation the IR singularities are then translated to
photon mass regularization \cite{Dittmaier:2003bc}.



\subsection{Real corrections}

\label{sect-real}

In addition to the virtual corrections we also have to consider
real photon emission, {\it i.e.} the processes $\eewwzgam$ and
$\eezzzgam$. The corresponding amplitudes are divergent in the
soft and collinear limits. The soft singularities cancel against
the ones in the virtual corrections while the collinear
singularities are regularized by the physical electron mass.  To
extract the singularities from the real corrections and combine
them with the virtual contribution we apply both the dipole
subtraction scheme and a phase-space slicing method.

%
%

The dipole subtraction formalism is a process independent approach and was
originally introduced for massless QCD \cite{Catani:2000ef}. We use a
generalization of this method to include photon radiation from massive
fermions \cite{Dittmaier:1999mb}. Since photon radiation from a massive
fermion and from a massive charged gauge boson has the same singular
structure in the soft limit this formalism can directly be applied to our
calculation.

In the dipole subtraction method a specially constructed function is
subtracted from the real amplitude and then added back
\begin{equation}
  \label{eq:dipmaster}
  \sigma_{\text{real}} = \int_4 \left( \dsreal - \dssub \right)
  + \int_4 \dssub.
\end{equation}
The subscript $4$ refers to the 4-body final state including
photon radiation. The subtraction
function has the same singular structure as the amplitude
pointwise in phase space. The difference of the real amplitude
and the subtraction function is therefore regular and the
integration of the first term in \eq{eq:dipmaster} can be
performed numerically. Introducing a photon mass as a regulator
the subtraction function can be integrated analytically over the
photon emission phase space up to a convolution
\begin{equation}
  \label{eq:dipsubint}
  \int_4 \dssub =- \frac{\alpha}{2\pi} \int \text{d}x \sum_{i \neq j}
  Q_i Q_j \, {\cal G}_{ij}(x) \int_3 \dsborn
  + \sigma_{\text{endpoint}},
\end{equation}
where the charges $Q_i$ are counted as outgoing and the subscript 3 refers
to the $3$-body phase space without photon radiation.
While the first term in
\eq{eq:dipsubint} has only mass singularities, the endpoint contribution
contains both soft and collinear singularities and is given by
\begin{equation}
  \label{eq:dipep}
  \sigma_{\text{endpoint}} = - \frac{\alpha}{2\pi} \int_3
  \dsborn \sum_{i \neq j} Q_i Q_j \, G_{ij}.
\end{equation}
The summations in \eq{eq:dipsubint} and \eq{eq:dipep} run over all
initial and final state charged particles. The explicit
expressions for the $G_{ij}$ and the ${\cal G}_{ij}(x)$ can be
found in \cite{Dittmaier:1999mb}. The soft and collinear
singularities cancel in the sum of the endpoint and virtual
contributions. The mass singularities in the final result then
originate only from the first term in \eq{eq:dipsubint}.

%
%

The idea of phase-space slicing is to split the phase space of the photon
emission contribution into a soft, a collinear and a remaining finite
part. In the soft and collinear regions the real amplitude approximately
factorizes into universal soft and collinear functions and the Born
amplitude. In addition the phase space splits into the leading order phase
space and a soft or collinear part. The phase-space integration over the
photon degrees of freedom can then be performed analytically resulting in
infrared and mass singular contributions. In the remaining part of
phase space the amplitude is regular and the integration can be performed
using numerical integration.

We have implemented a two-cutoff phase-space slicing method closely
following \cite{Baur:1998kt, Denner:2000bj}. The
differential cross section for the real contribution is decomposed
as follows
\begin{eqnarray}
d\sigma_{real} &=& d\sigma_{soft}(\delta_s) + d\sigma_{hard}(\delta_s),\crn
d\sigma_{hard}(\delta_s) &=& d\sigma_{coll}(\delta_s,\delta_c) + d\sigma_{fin}(\delta_s, \delta_c)
\end{eqnarray}
using the soft and collinear cutoffs $\delta_s$ and $\delta_c$. The soft
region is defined by $E_{\gamma}<\delta_s \sqrt{s}/2=\Delta E$.  The
collinear but non-soft region is given by \{$E_{\gamma}\ge \Delta E$, $1 -
\cos\theta_{\gamma f} < \delta_c$\} where $\theta_{\gamma f}$ is the polar
angle of the photon with respect to the $e^\pm$ direction in the
c.m. frame.

\begin{figure}[t]
  \centering
  \includegraphics[width=0.6\textwidth]{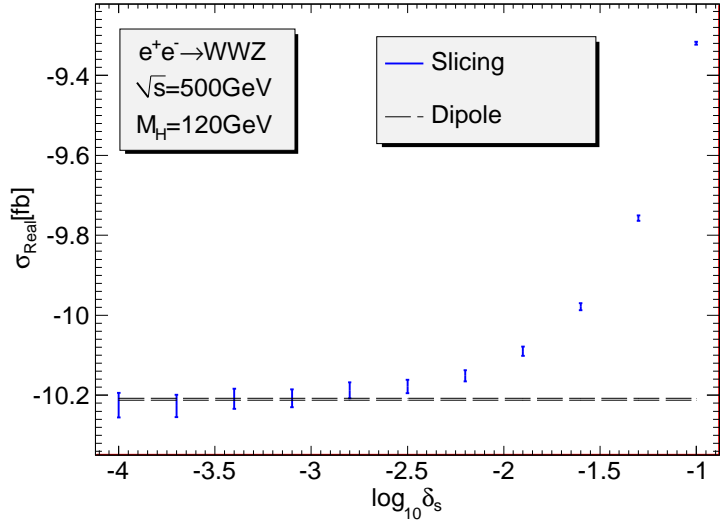}
  \caption{{\em Dependence of $\sigma_{\text{real}}^{\eewwzgam}$ on the soft
      cutoff $\delta_s$ in phase-space slicing with fixed $\delta_c = 7
      \cdot 10^{-4}$. Only the non-singular part is shown, i.e. the IR singular
      $\ln(m_\gamma^2)$ terms are set to zero. The result using dipole
      subtraction is shown for comparison with the error given by the width
      of the band.}}
  \label{fig:dipslicmp}
\end{figure}

Since the approximations used in the soft and collinear regions introduce
errors of ${\cal O}(\delta_s, \delta_c)$ the cutoffs have to be chosen
sufficiently small.  As can be seen from \fig{fig:dipslicmp} the
approximation errors are below the integration errors for $\delta_s \leq
10^{-3}$. For smaller values of the soft cutoff the errors grow larger due
to cancellations between the soft and the hard contributions which both
diverge as $\log\delta_s$.  We have similarly verified the stability with
respect to the variation of the collinear cutoff $\delta_c$.
\fig{fig:dipslicmp} also shows agreement between the slicing and dipole
subtraction results, although the errors are typically a factor 10 smaller
when using the subtraction formalism.



\subsection{Defining the  weak corrections}
It is well known that the collinear QED correction related to
initial state radiation (ISR) in $e^{+}e^{-}$ processes is large.
The effect due to ISR can be treated along a structure function
approach which resums the effects of higher orders, see for
example \cite{Boudjema:1996qg}. Likewise for the linear collider
the effect of beamstrahlung \cite{Barklow:1992ci} needs to be
convoluted over the genuine weak corrections. Therefore in a NLO
computation such as ours it is best to subtract the ISR
corrections in order to sum their effect to all orders or, put
another way, once the weak correction has been defined to
convolute its result within a structure function
approach. Deconvolution is also possible
from the experimental data to arrive at the non-ISR result as is
done at LEP, for example \cite{Boudjema:1996qg}. The weak corrections encapsulate more
interesting features. In order to see the effect of the weak
corrections, one should separate the large QED corrections from
the full NLO result. It means that we can define the weak
correction as an infrared and collinear finite quantity. In our
work we will subtract all of the QED corrections, not only the
initial but, for $WWZ$, also the final and the interference QED
corrections. The definition we adopt in this paper is based on the
dipole subtraction formalism. In this approach, the sum of the
virtual and endpoint contributions satisfies the above
conditions\footnote{A similar discussion was given in
\cite{Denner:2000bj}.} and can be chosen as a definition for the
weak correction
\begin{eqnarray}
\sigma_{\text{weak}} = \sigma_{\text{virt}} + \sigma_{\text{endpoint}}.
\end{eqnarray}
For the numerical results shown in \sect{sect-eezzz} and \sect{sect-eewwz},
we will make use of this definition.

If one uses the phase-space slicing approach, a definition for the
weak correction can be found as well. In the neutral process
\eezzzt the QED corrections are confined to the initial state. The
universal leading QED part of $\sigma_{\text{virt}} +
\sigma_{\text{soft}}$ can be extracted and is given by\footnote{This
definition differs by the sub-sub-leading term  $2 \alpha/\pi
\times \pi^2/6$ to what is usually taken for the universal initial
state radiation. Adding this term would give the same result as
the one based on a diagrammatic approach as done in
\cite{JiJuan:2008nn} for $ZZZ$.}
\beqn
\label{dqeduniv} \sigma_{V+S}^{\text{QED}}=\frac{2
\alpha}{\pi}\left((L_e-1)\ln \delta_s +\frac{3}{4}L_e
-1 \right) \sigma_{\text{Born}} \;,\; L_e=\ln(s/m_e^2) \;.
\eeqn
Subtracting this from the sum of virtual and soft contributions we can
define the weak corrections in phase-space slicing as
\begin{equation}
  \sigma_{\text{weak, slicing}}^{ZZZ} = \sigma_{\text{virt}} + \sigma_{\text{soft}} - \sigma_{V+S}^{\text{QED}}.
\end{equation}
This procedure will lead to the same result for the weak correction
as obtained by simply taking the sum of the virtual and endpoint parts.

For \eewwzt there is also final state radiation and its
interference with the initial state radiation. Diagrammatically
there is no unambiguous way to subtract this contribution in a
gauge invariant way. Nonetheless after subtracting the initial
state radiation in \eq{dqeduniv}, the logarithms of infrared
origin can be easily isolated and the weak correction defined by
\begin{equation}
  \sigma_{\text{virt}} + \sigma_{\text{soft}} - \sigma_{V+S}^{\text{QED}} =
  \sigma_{\text{weak, slicing}}^{WWZ} + b \ln\delta_s.
\end{equation}
The coefficient $b$ of $\ln \delta_s$ can be extracted by choosing
two different values of $\delta_s$ (which should be sufficiently
small).  We have compared the weak correction obtained in this way
with the one calculated by taking the sum of the virtual and
endpoint parts. The results for the case of $\sqrt{s}=500\gev$ and
$M_H=120\gev$ are: $\delta_{weak}^{dipole}=-7.014(5)\%$,
$\delta_{weak}^{slicing}= -6.73(1)\%$. This is for the process
$\eewwz$ and other input parameters as specified in
\sect{sect-input-wwz}.



\subsection{Input parameters}
\label{sect-input-wwz}
We use the following set of input parameters \cite{Amsler:2008zzb,
:2009ec}, \begin{equation}
\begin{aligned}
G_\mu &= 1.16637\times 10^{-5}\gev^{-2}, \hs \alpha(0)=1/137.035999679 , \\
M_{W} &= 80.398\gev, \hs M_Z= 91.1876\gev,  \\
m_e &= 0.510998910\mev, \hs m_\mu = 105.658367\mev, \hs m_\tau = 1776.84\mev ,\\
m_u &= 66\mev, \hs m_c = 1.2\gev, \hs m_t=173.1\gev, \\
m_d &= 66\mev, \hs m_s = 150\mev, \hs m_b = 4.3\gev.
\end{aligned}
\end{equation} The masses of the light quarks\footnote{which are the same as
those used in \cite{Bredenstein:2006rh}.}, {\it i.e.} all but the top mass,
are effective parameters
adjusted to reproduce the hadronic contribution to the photonic
vacuum polarization of \cite{Jegerlehner:2001ca} with
$\alpha^{-1}(M_Z^2)=128.907$. As discussed in \ssect{sec:calc-ren}
we use a variant of the $G_\mu$ scheme with $\alpha_{G_\mu}$ at
leading order leading to NLO corrections that are of ${\cal
O}(\alpha_{G_\mu}^3\alpha(0))$. Using $\alpha_{G_\mu}$ as coupling
we calculate  $\Delta r=3.0792\times 10^{-2}$ for $M_H=120\gev$
and $\Delta r=3.1577\times 10^{-2}$ for $M_H=150\gev$. The
Cabibbo-Kobayashi-Maskawa matrix is set to be diagonal. For the
calculation we neglect the electron Yukawa coupling proportional
to the electron mass, as mentioned at the beginning of
\sect{sect-code}. For both processes we apply no cuts at the level
of the $W^\pm$ and $Z$, since these will decay.


\section{$\eezzz$}
\label{sect-eezzz}
\begin{figure}[t]
\begin{center}
\mbox{\includegraphics[width=0.47\textwidth]{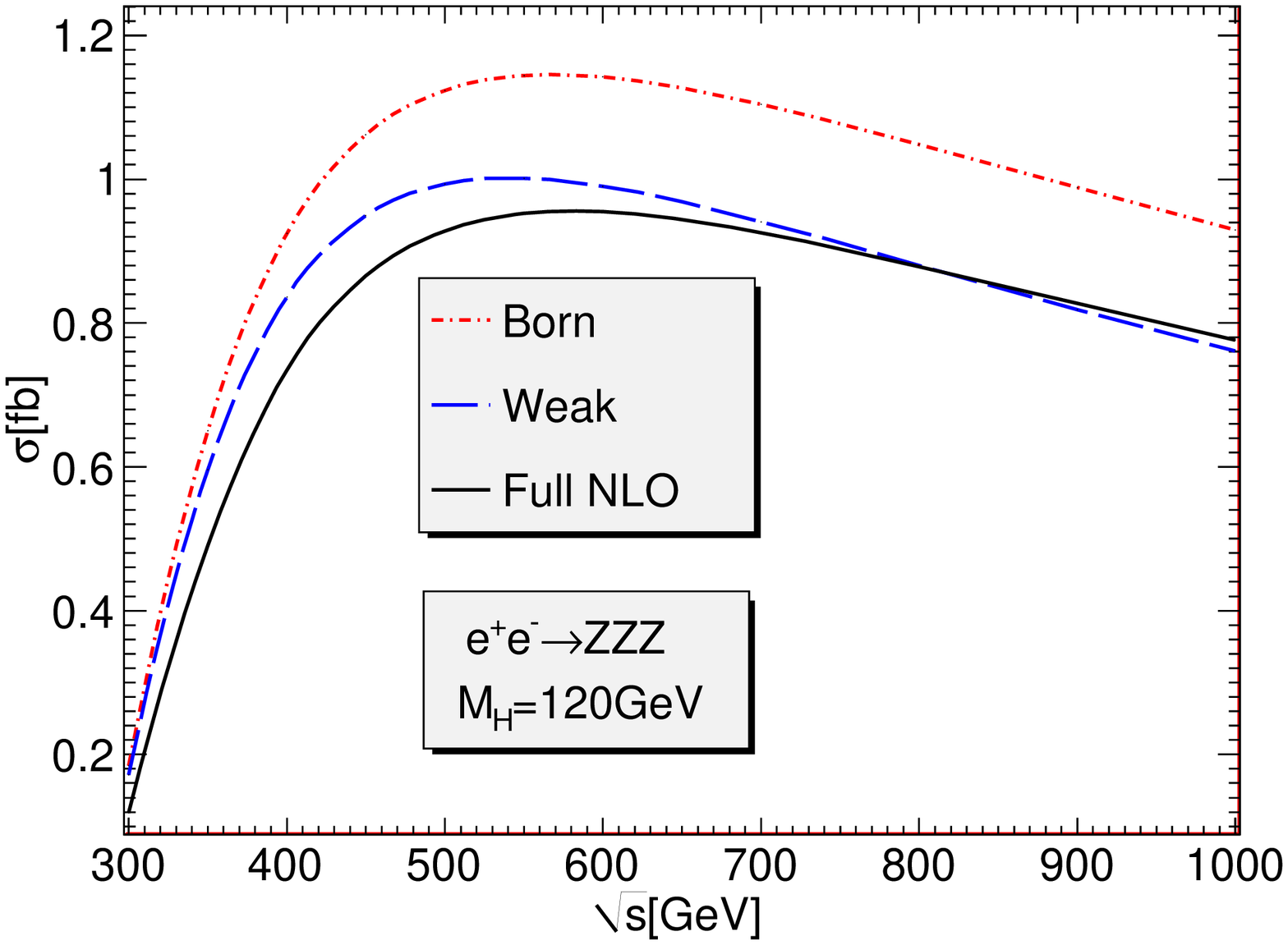}
\hspace*{0.01\textwidth}
\includegraphics[width=0.47\textwidth]{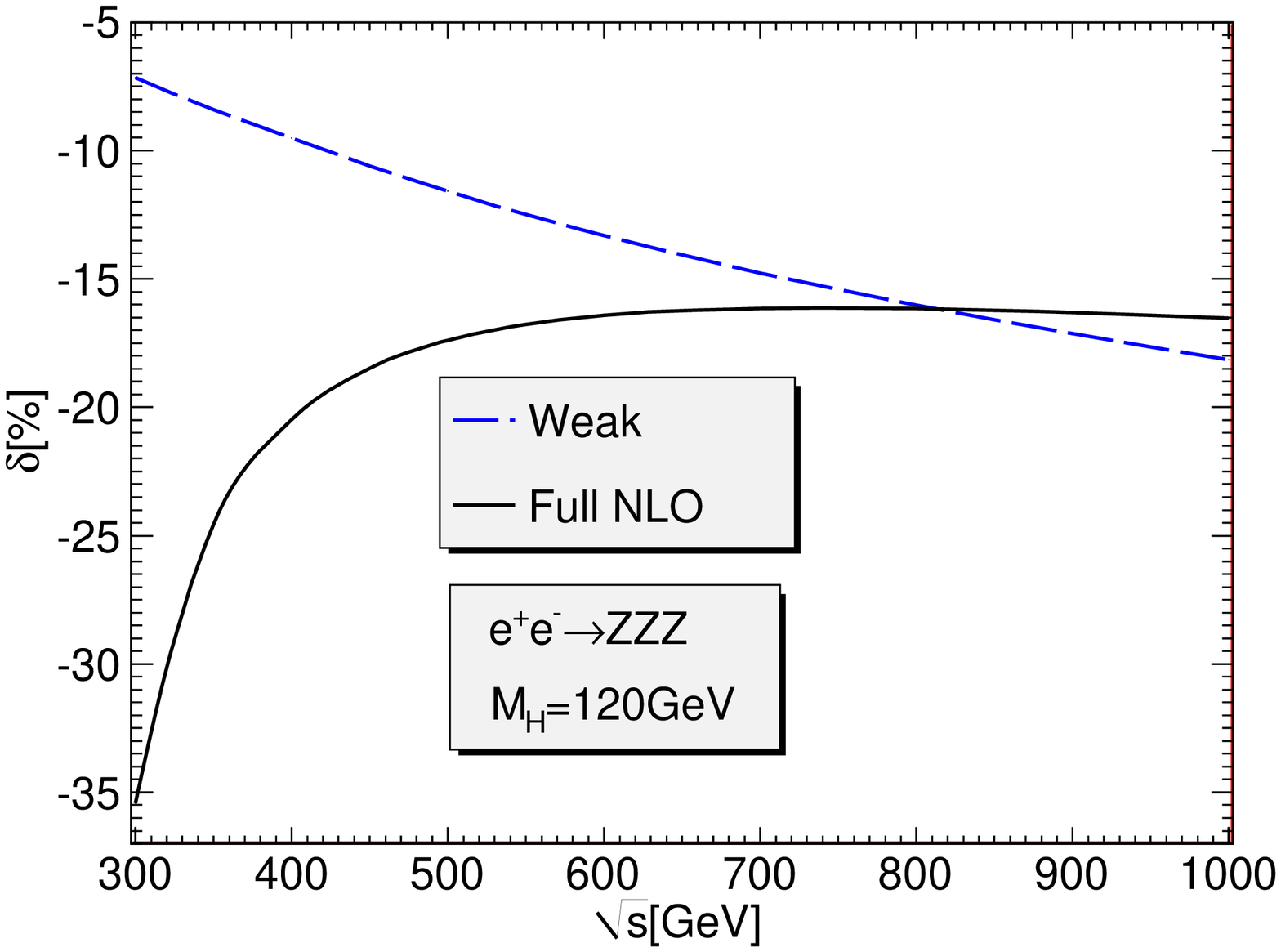}}
\caption{\label{ourzzzplots}{\em Left: the total cross section
for $\eezzz$ as a function of $\sqrt{s}$ for the Born, full ${\cal
O}(\alpha)$
 and genuine weak correction for $M_H=120\gev$. Right: the
corresponding relative percentage corrections
$\sigma_{NLO}/\sigma_{LO}-1$.}}
\end{center}
\end{figure}
As shown in Fig.~\ref{ourzzzplots} the tree-level cross section
rises sharply once the threshold for production opens, reaches a
peak of about $1.1\fb$ around a centre-of-mass energy of $600\gev$
before very slowly decreasing with a value of about $0.9\fb$ at
1\tev. Exact results are displayed in Table~\ref{tab-eezzz-rs}.
The Higgsstrahlung contribution to the total cross
section is  about $10\%$  at $\sqrt{s}=600\gev$ and
$\sqrt{s}=1\tev$.

The full NLO corrections are quite large and negative around
threshold, $-35\%$, decreasing sharply to stabilise at a plateau
around $\sqrt{s}=600\gev$ with $-16\%$ correction. The sharp rise
and negative corrections at low energies are easily understood.
They are essentially due to initial state radiation (ISR) and the
behaviour of the tree-level cross section. The photon radiation
reduces the effective centre-of-mass energy and therefore explains
what is observed in the figure. On the other hand the genuine weak
corrections, in the $G_\mu$ scheme, are relatively small at
threshold, $-7\%$. The magnitude of the corrections however
increases steadily reaching a value as large as $-18\%$ at
$\sqrt{s}=1\tev$. These large negative corrections are typical of
 the electroweak Sudakov logarithms $-\log^2 (s/M_W^2)$. In the
usual $\alpha(0)$ on-shell scheme this important effect would be
blurred and weakened unless one reaches much higher energies.
\begin{table}[t]
\small \bc \caption{\em Cross section for $\eezzz$ at
tree-level, including only the weak corrections and at full next-to-leading
order for $M_H=120\gev$. Also shown are the relative weak and full NLO corrections.}
\vspace*{1ex}
\begin{tabular}{c r@{.}l r@{.}l r@{.}l r@{.}l r@{.}l}
    \hline
$\sqrt{s}\,[\tev]$
& \multicolumn{2}{c}{$\sigma_{Born}[\fb]$}
& \multicolumn{2}{c}{$\sigma_{weak}[\fb]$}
& \multicolumn{2}{c}{$\sigma_{full}[\fb]$}
& \multicolumn{2}{c}{$\delta_{weak}[\%]$}
& \multicolumn{2}{c}{$\delta_{full}[\%]$}
\\ \hline
0.3 & 0&184524(4) &0&17132(1)& 0&11916(2) & -7&156(7)& -35&420(8) \\
    \hline
0.4 & 0&92437(3) &0&83637(9)& 0&73502(9) &-9&520(9)& -20&484(9) \\
    \hline
0.5 & 1&12353(1) &0&99338(9)& 0&92820(9) &-11&584(7)& -17&386(8)  \\
    \hline
0.6 & 1&14203(6) &0&9900(2)& 0&9546(2)  &-13&31(1)& -16&41(1)  \\
    \hline
0.8 & 1&04796(8) &0&8801(2)& 0&8786(2)  & -16&02(2)& -16&16(2) \\
    \hline
1.0 & 0&92962(10) &0&7609(2)& 0&7759(2) &-18&15(2)& -16&54(2) \\
    \hline
\end{tabular}
\label{tab-eezzz-rs} \ec
\end{table}

\begin{figure}[htbp]
\begin{center}
\mbox{\includegraphics[width=0.45\textwidth
]{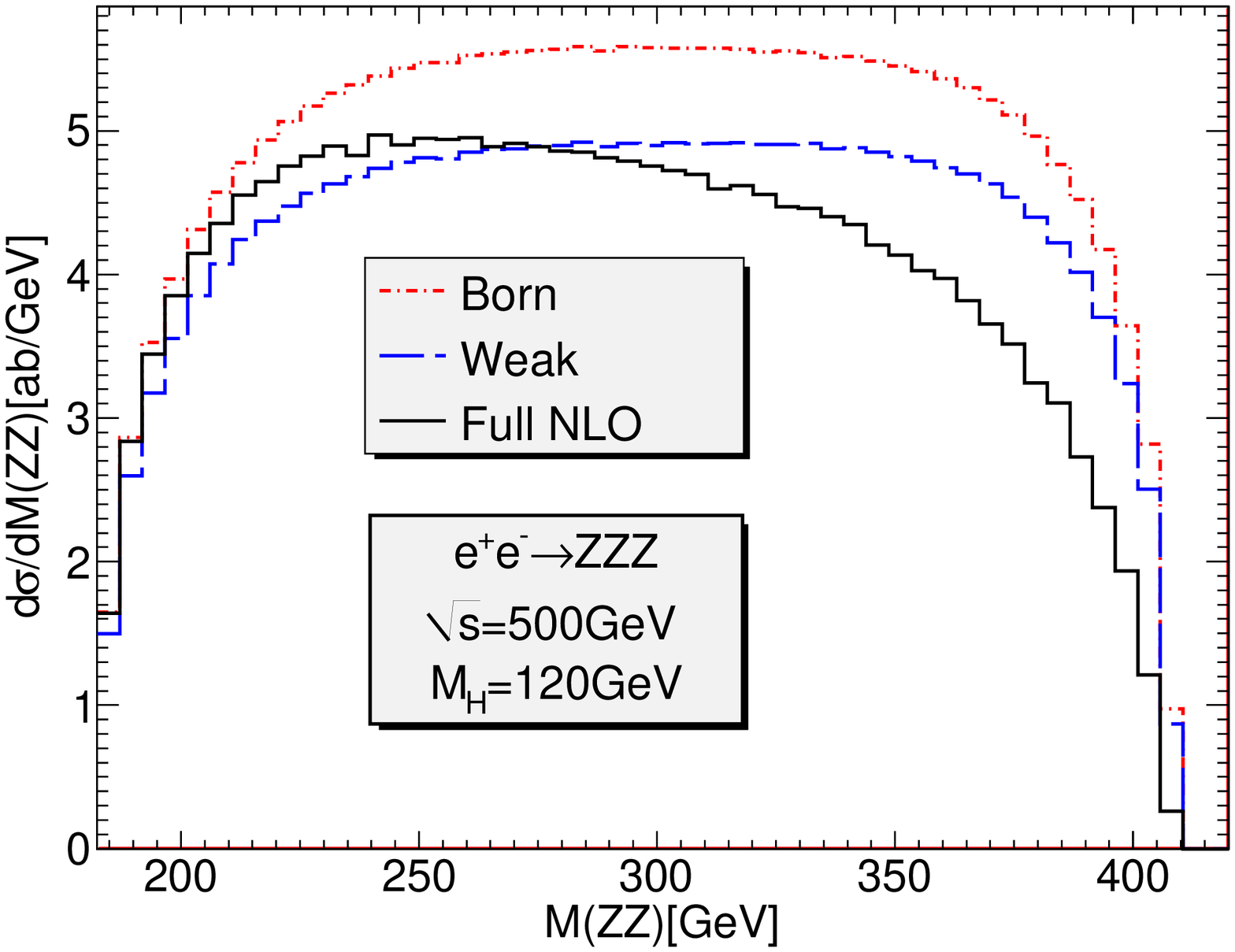} \hspace*{0.01\textwidth}
\includegraphics[width=0.45\textwidth]{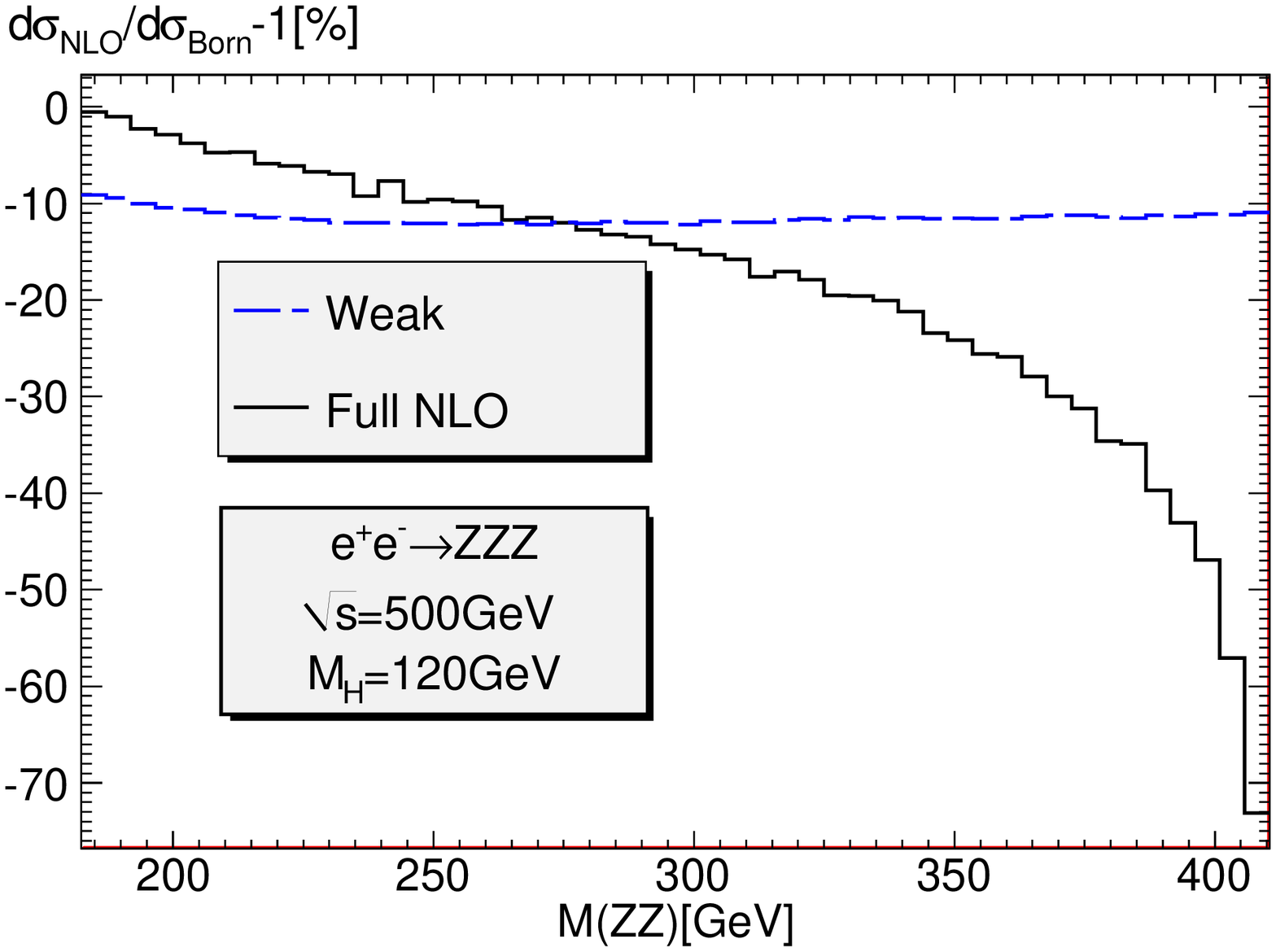}}
\mbox{\includegraphics[width=0.45\textwidth
]{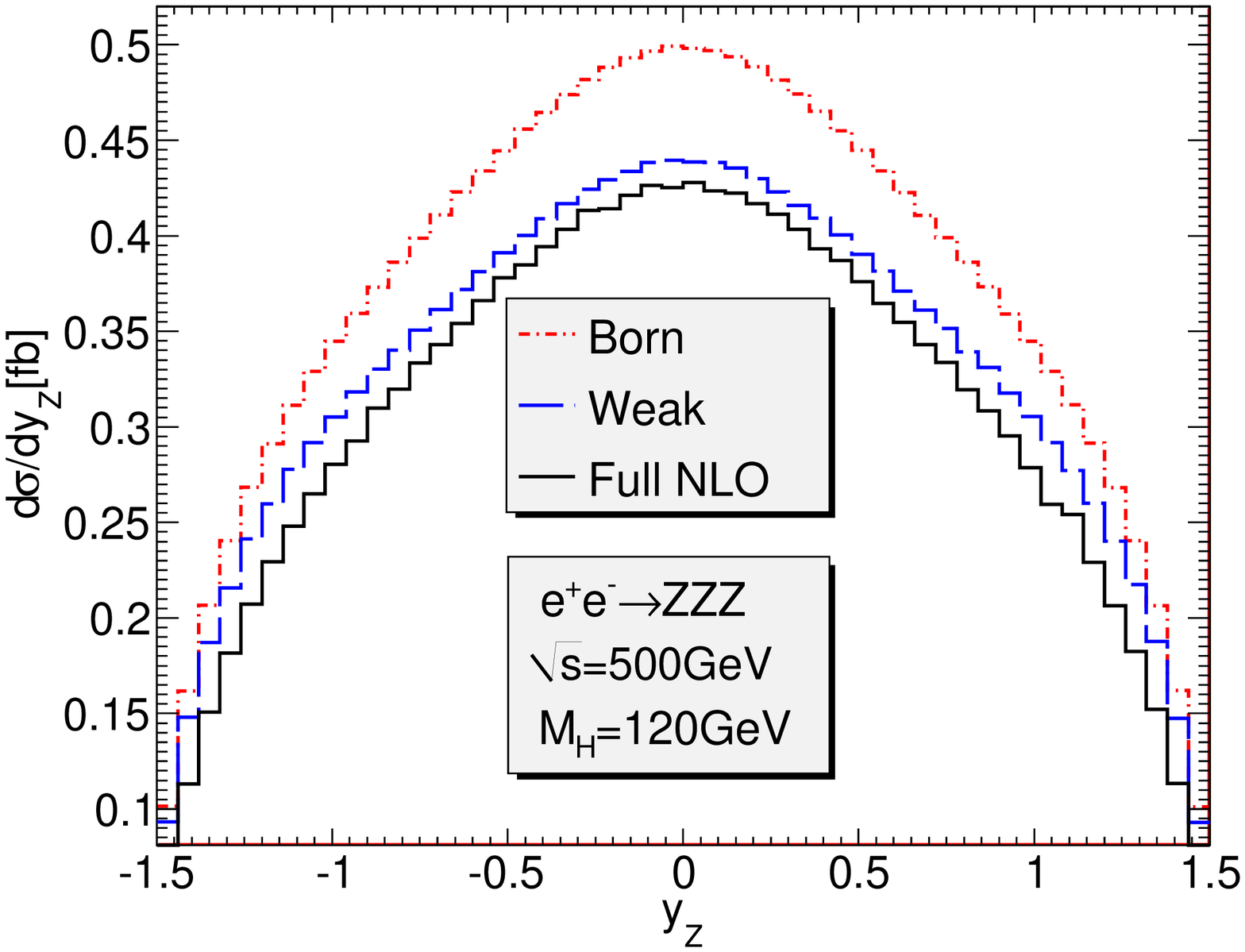} \hspace*{0.01\textwidth}
\includegraphics[width=0.45\textwidth]{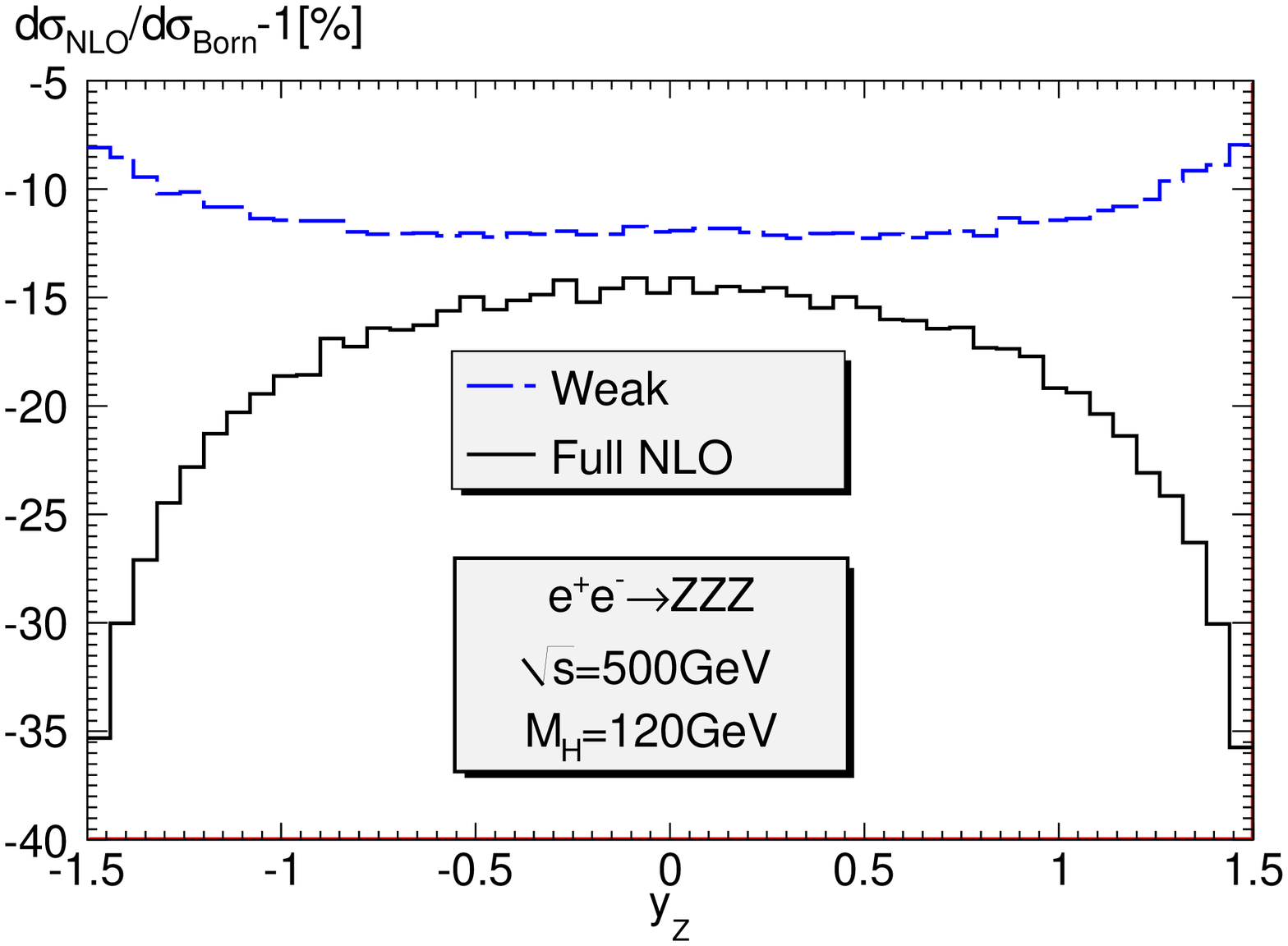}}
\mbox{\includegraphics[width=0.45\textwidth
]{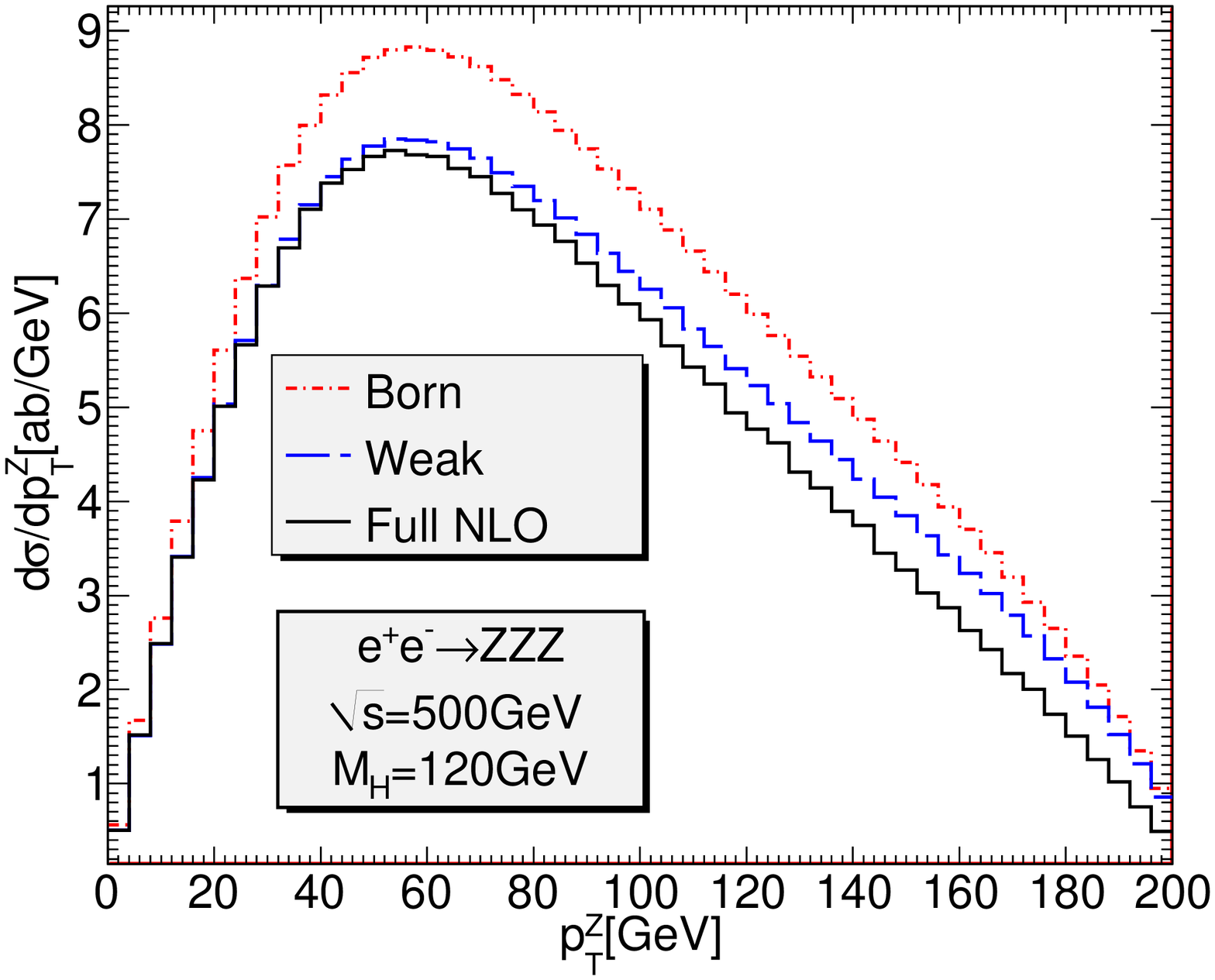} \hspace*{0.01\textwidth}
\includegraphics[width=0.45\textwidth]{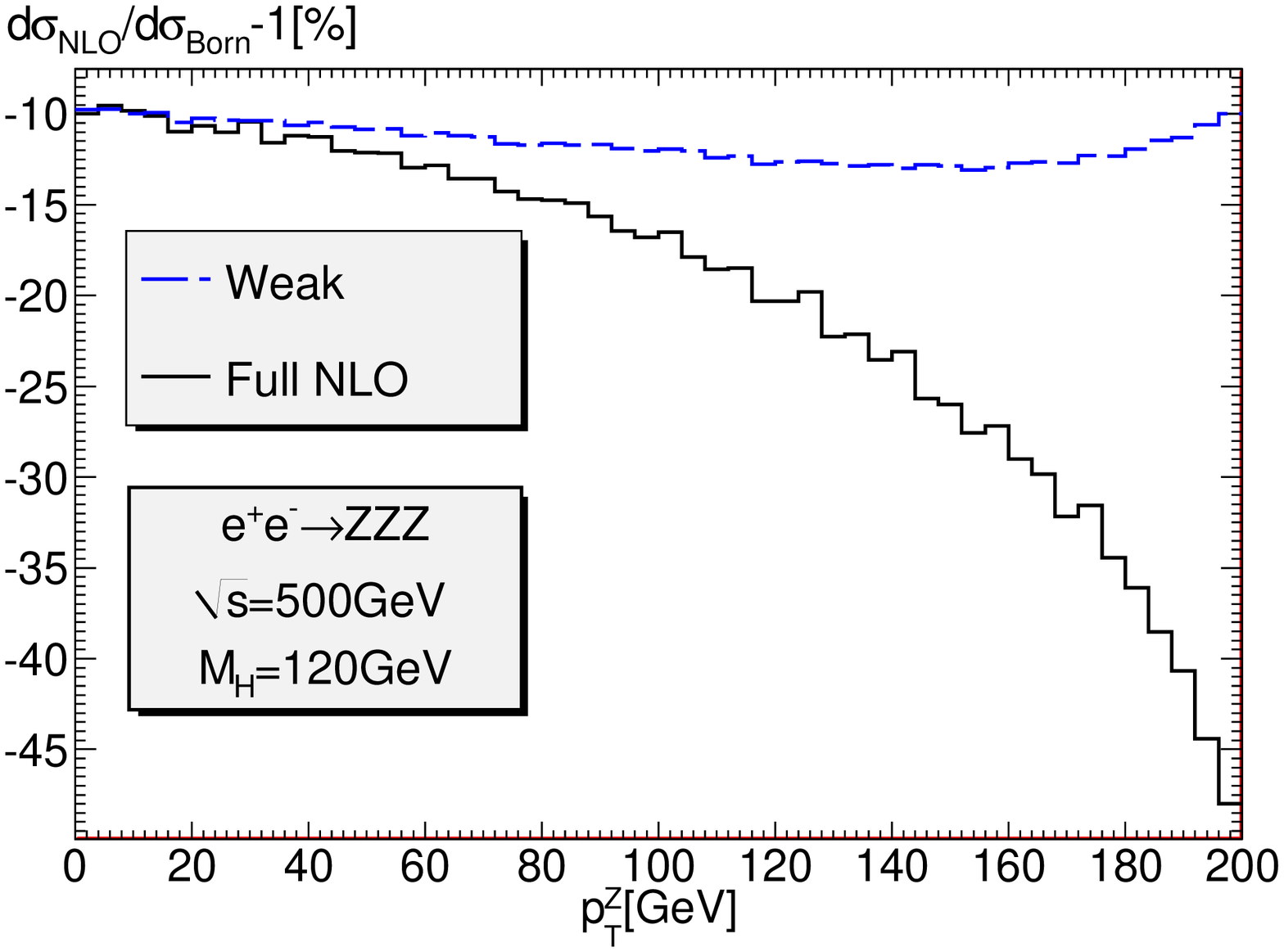}}
\caption{\label{ourzzzdist}{\em From top to bottom: distributions
for the $ZZ$ invariant mass, the rapidity of the $Z$  and the
transverse momentum of the $Z$ for \eezzzt at $\sqrt{s}=500\gev$
and $M_H=120\gev$. The panels on the left show the
tree-level, the full NLO and the
weak correction.
The panels on the right show the corresponding relative (to
the tree-level) percentage corrections. The distributions are
obtained by entering for each event the corresponding observable,
say $p_T^Z$, of each $Z$ and then normalising by a factor $1/3$.
}}
\end{center}
\end{figure}
We have also studied distributions in some key kinematic variables
and how they are affected by radiative corrections. Fig.~\ref{ourzzzdist} shows
the invariant mass $M_{ZZ}$, the transverse momentum $p_T^Z$ and the
rapidity $y_Z$ for $\sqrt{s}=500\gev$ and $M_H=120\gev$. The
important message is that the genuine weak corrections are almost
 just an overall rescaling of the leading-order distributions,
in particular in the bulk of the region of phase space
where the yield is largest. The full NLO corrections on the other
hand show more structure with very large corrections at the edges
of phase space where the cross section is smallest, for
example the full NLO correction for $M_{ZZ}>350\gev$ drops below
$-50\%$. This again is essentially due to hard photon radiation.
This shows that the effects of New Physics could be discovered in
a less ambiguous way after subtracting the QED corrections.

\section{$\eewwz$}
\label{sect-eewwz}
\begin{figure}[t]
\begin{center}
\mbox{\includegraphics[width=0.48\textwidth,height=0.4\textwidth
]{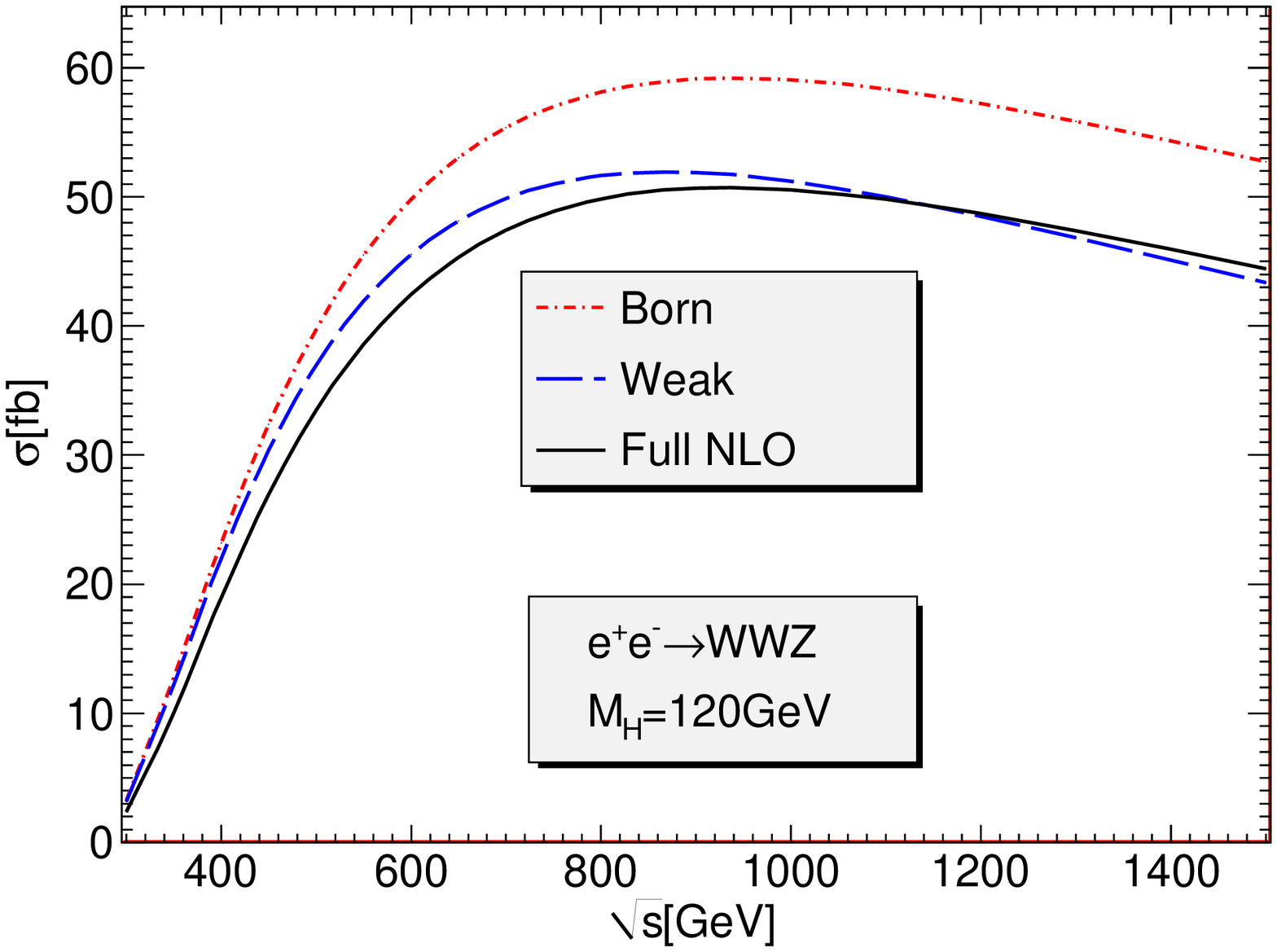} \hspace*{0.01\textwidth}
\includegraphics[width=0.48\textwidth,height=0.4\textwidth]{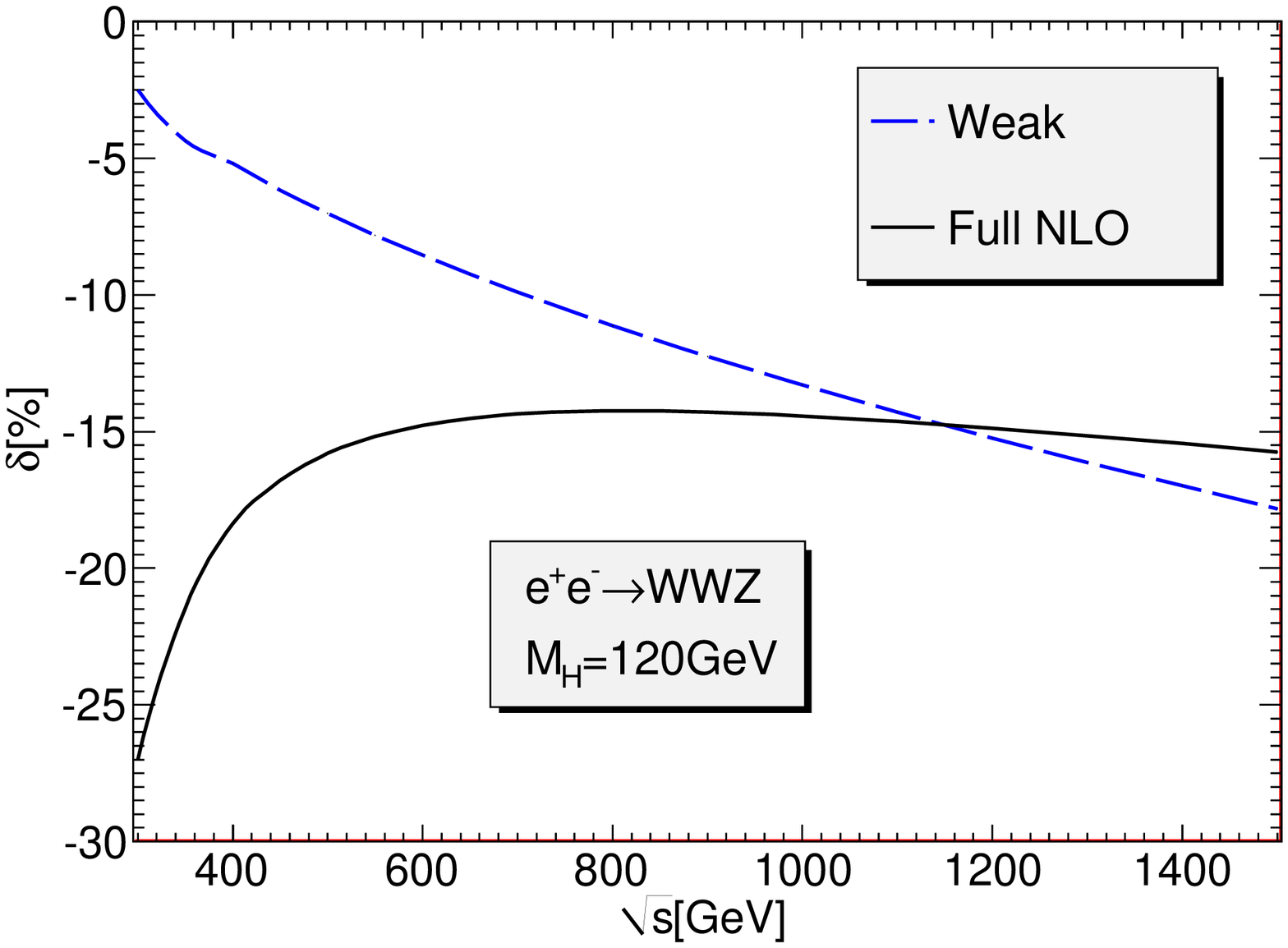}}
\caption{\label{ourwwzplots}{\em Left: the total cross section for $\eewwz$
as a function of $\sqrt{s}$ for the Born, full ${\cal O}(\alpha)$
 and genuine weak correction for $M_H=120\gev$. Right: the
corresponding relative percentage corrections
$\sigma_{NLO}/\sigma_{LO}-1$.}}
\end{center}
\end{figure}

Compared to  \eezzztp, the cross section for
\eewwzt is almost 2 orders of magnitudes larger for the same
centre-of-mass energy. For example at $500\gev$ it is about $40\fb$ at tree
level,
compared to $1\fb$ for the \eezzzt cross section. For an anticipated
 luminosity of $1 {\rm ab}^{-1}$, this means that
the cross section should be known at the
per-mil level.
 In absolute terms the
Higgsstrahlung contribution is a factor 2 (due to $SU(2)$) larger
than in \eezzztp, however its contribution to the total \eewwzt
cross section is much less than $1\%$ for $M_H=120\gev$.

\begin{table}[h]
\small \bc \caption{\em Cross section for $\eewwz$ at
tree-level, including only the weak corrections and at full next-to-leading
order for $M_H=120\gev$. Also shown are the relative weak and full NLO corrections.}
\vspace*{1ex}
\begin{tabular}{c r@{.}l r@{.}l r@{.}l r@{.}l r@{.}l}
    \hline
$\sqrt{s}\,[\tev]$
& \multicolumn{2}{c}{$\sigma_{Born}[\fb]$}
& \multicolumn{2}{c}{$\sigma_{weak}[\fb]$}
& \multicolumn{2}{c}{$\sigma_{full}[\fb]$}
& \multicolumn{2}{c}{$\delta_{weak}[\%]$}
& \multicolumn{2}{c}{$\delta_{full}[\%]$}
\\ \hline
0.3 & 3&27055(4) & 3&1888(3) & 2&3880(3) &-2&500(8)&-26&986(9)  \\
    \hline
0.5 & 39&7557(9) & 36&967(2) & 33&476(2) &-7&014(5)&-15&795(5)  \\
    \hline
0.7 & 55&358(3) & 49&878(6) & 47&409(6) &-9&899(10)&-14&359(10)  \\
    \hline
0.9 & 59&121(4) & 51&881(8) & 50&678(8) &-12&25(1)&-14&28(1)  \\
    \hline
1.0 & 59&061(4) & 51&206(9) & 50&541(9) &-13&30(1)&-14&43(1)  \\
    \hline
1.2 & 57&202(5) & 48&49(1) & 48&69(1) &-15&24(2)&-14&88(2)  \\
    \hline
1.5 & 52&740(5) & 43&34(1) & 44&43(1) &-17&82(2)&-15&76(2)  \\
    \hline
\end{tabular}
\label{tab-eewwz-rs} \ec
\end{table}

The behaviour of  the \eewwzt cross section as a function of
energy resembles that of \eezzztp. It rises sharply once the
threshold for production opens, reaches a peak before very slowly
decreasing as shown in Fig.~\ref{ourwwzplots} (exact results
are also displayed in Table~\ref{tab-eewwz-rs}). However as already
discussed the value of the peak is much larger, $\sim 50\fb$ at NLO,
moreover the peak is reached around $\sqrt{s}=1\tev$, much higher than
in $ZZZ$. This explains the bulk of the NLO corrections at lower
energies which are dominated by the QED correction, large and
negative around threshold and smaller at higher energies. As
the energy increases the weak corrections get larger
reaching about $-18\%$ at $\sqrt{s}=1.5\tev$. Again a
large part of this correction seems to be of the Sudakov type.

\begin{figure}[htbp]
\begin{center}
\mbox{\includegraphics[width=0.45\textwidth
]{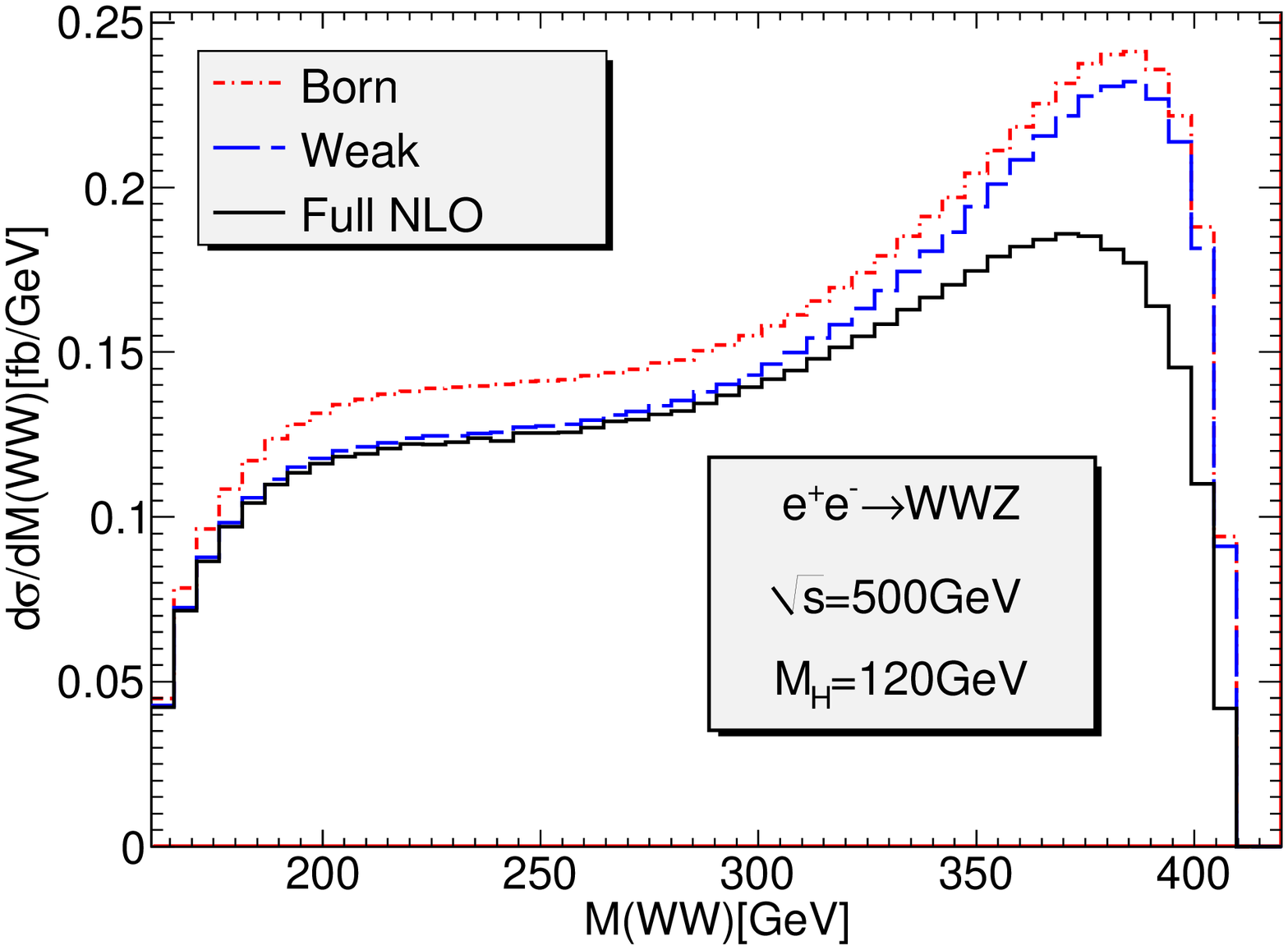} \hspace*{0.01\textwidth}
\includegraphics[width=0.45\textwidth]{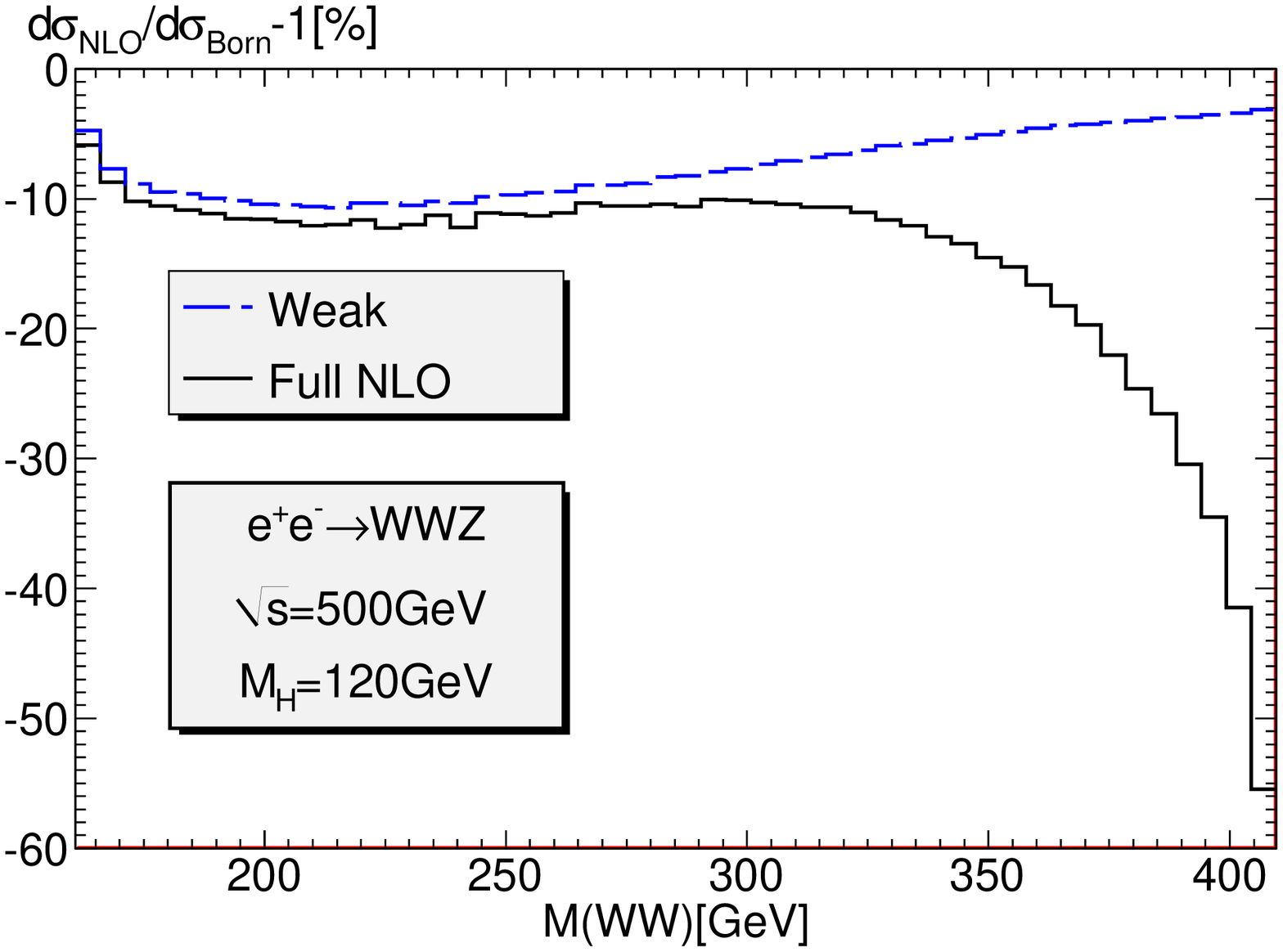}}
\mbox{\includegraphics[width=0.45\textwidth
]{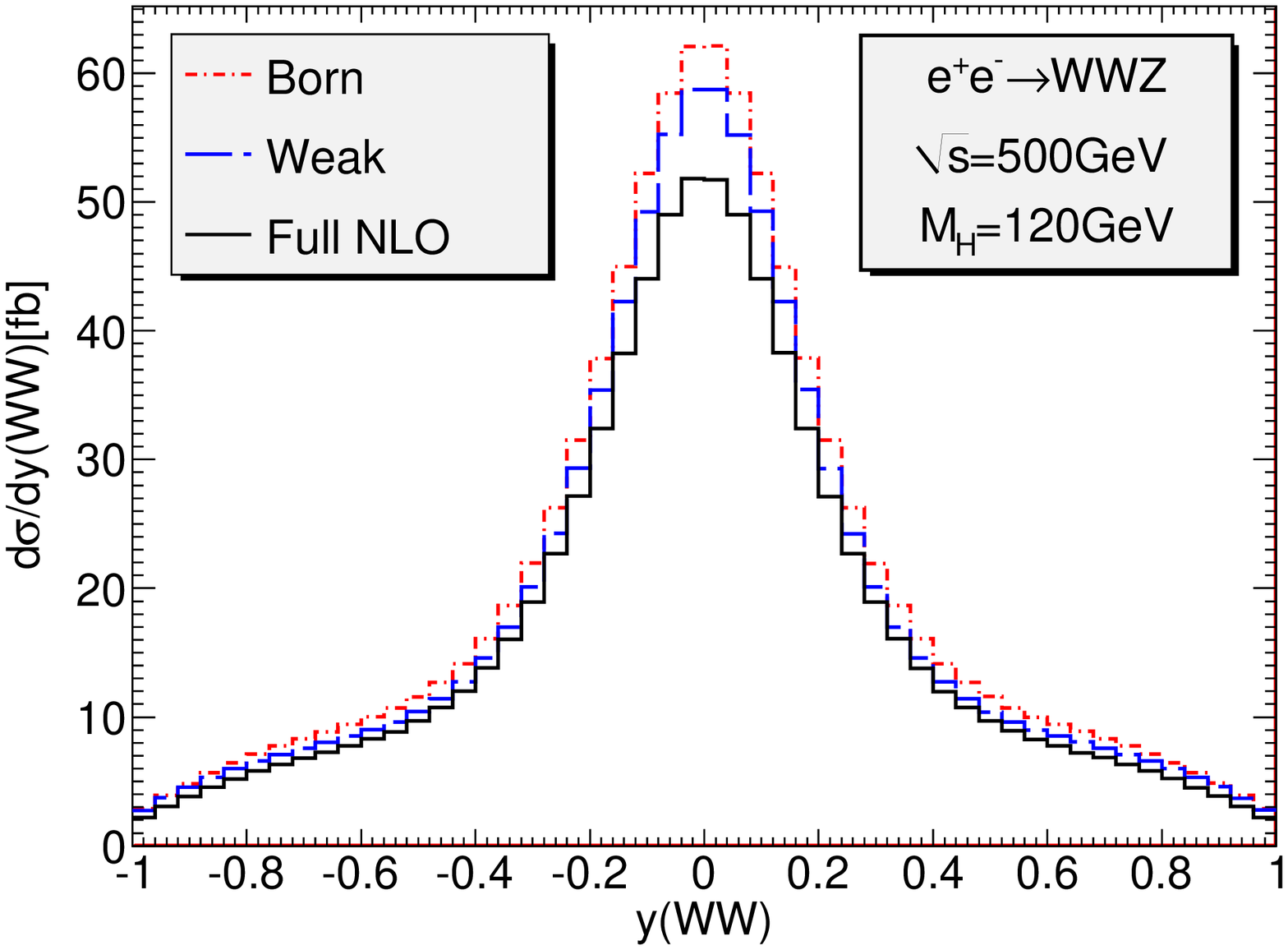} \hspace*{0.01\textwidth}
\includegraphics[width=0.45\textwidth]{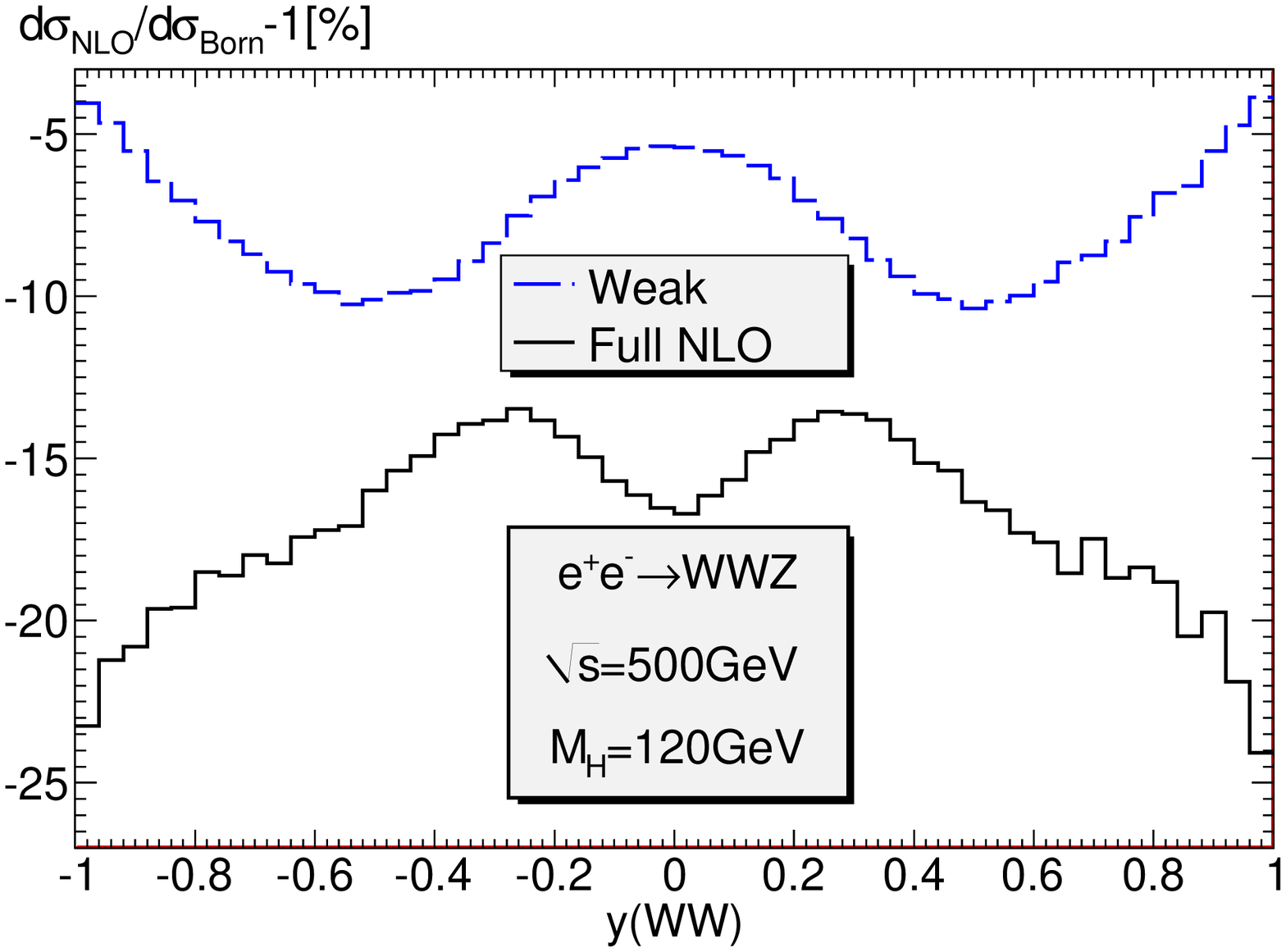}}
\mbox{\includegraphics[width=0.45\textwidth
]{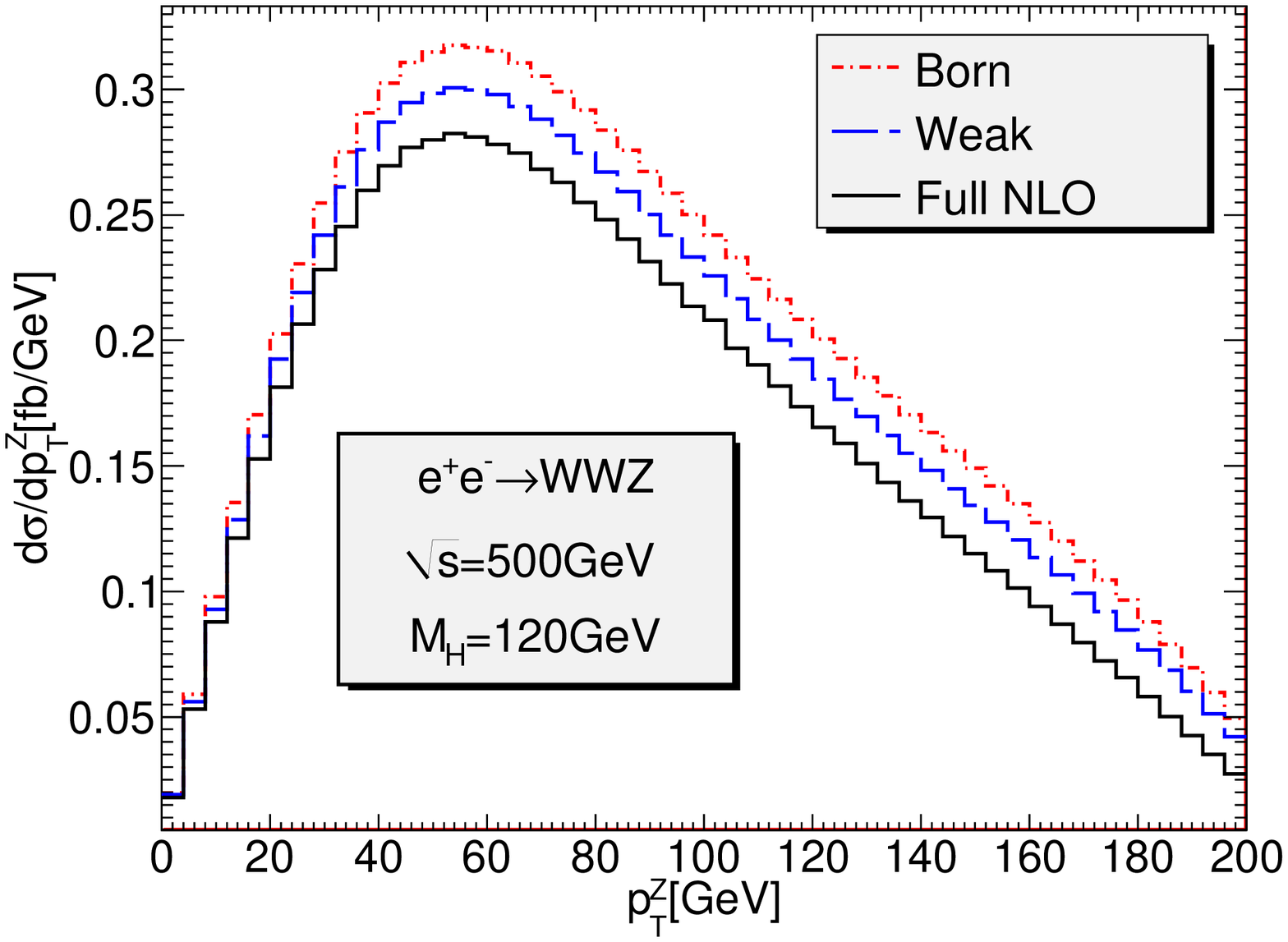} \hspace*{0.01\textwidth}
\includegraphics[width=0.45\textwidth]{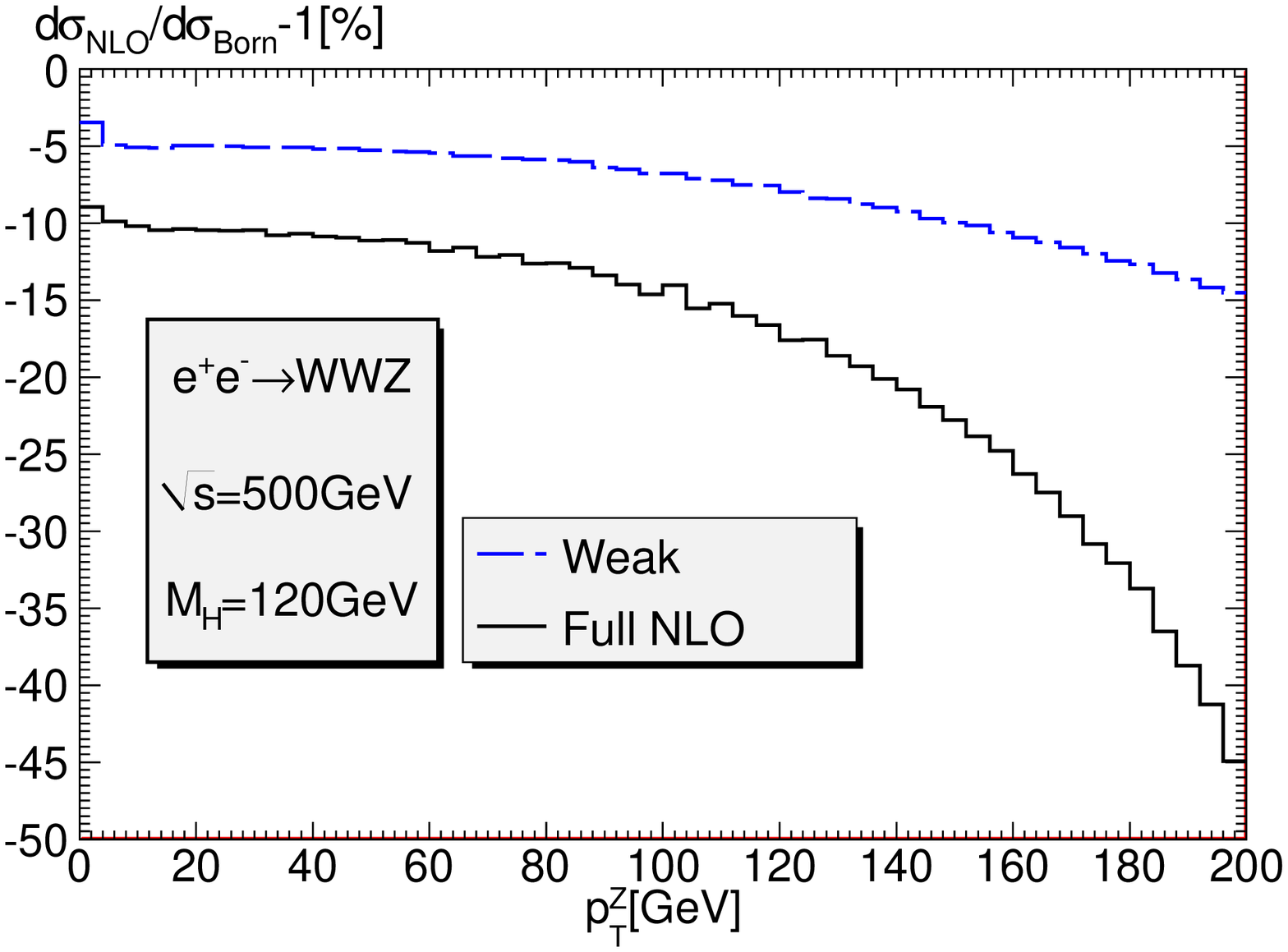}}
\caption{\label{ourwwzdist}{\em From top to bottom: distributions
for the $WW$ invariant mass, the rapidity of the $WW$ system and
the transverse momentum of the $Z$ for \eewwzt at
$\sqrt{s}=500\gev$ and $M_H = 120\gev$. The panels on
the left show the tree-level, the full NLO and the weak correction.
The panels on the right show the
corresponding relative (to the tree-level) percentage corrections.
}}
\end{center}
\end{figure}

More interesting than in the case of \eezzzt are the distributions
in some key variables like the invariant $WW$ mass, the $p_T^Z$
and the rapidity of the $WW$ system. First, due to photon
radiation, in the full NLO corrections some large corrections do
show up at the edges of phase space, see Fig.~\ref{ourwwzdist}.
However when the QED corrections are subtracted, the weak
corrections cannot be parameterised by an overall scale factor,
for all the distributions that we have studied. Other
distributions not shown here can be found in
\cite{Boudjema:2010ad}.

\section{Comparison with other calculations}
\label{sect-cmp-ma}

The electroweak corrections to $\eezzz$ have also been calculated
in \cite{JiJuan:2008nn}. They use the $\alpha(0)$ scheme and
slightly different input parameters. We have performed a tuned
comparison by adapting to the input parameters of
\cite{JiJuan:2008nn} and switching to the $\alpha(0)$ scheme. We
find full agreement within the quoted statistical errors for
$\sigma_{LO}$ and $\Delta\sigma_{tot}$ shown in Table~2 of
\cite{JiJuan:2008nn}. A comparison of the results in Fig.~4 of
\cite{JiJuan:2008nn} is shown in \tab{tab-zzz-compa}\footnote{The
data for
  Fig.~4 of \cite{JiJuan:2008nn} have been kindly provided by Prof.~Ma.}. The results for the
Born cross section agree to within $0.01\%$ while the NLO results
agree to at least $0.1\%$.
\begin{table}[h]
\small
\bc
\caption{\em Born cross section and relative corrections for
  $\eezzz$ using the input parameter scheme of
  \cite{JiJuan:2008nn}.}
\vspace*{1ex}
\begin{tabular}{cc@{\qquad}r@{.}l r@{.}l r@{.}l r@{.}l}
    \hline
\multicolumn{2}{c}{} & \multicolumn{4}{c}{$M_H=120\gev$} & \multicolumn{4}{c}{$M_H=150\gev$} \\
$\sqrt{s}\,[\gev]$ &  & \multicolumn{2}{c}{$\sigma_{Born}[\fb]$}
&\multicolumn{2}{c}{$\delta_{full} \,[\%]$} & \multicolumn{2}{c}{$\sigma_{Born}[\fb]$}
&\multicolumn{2}{c}{$\delta_{full} \,[\%]$} \\
    \hline
  350 & Ref.~\cite{JiJuan:2008nn} & 0&58696        &  -15&79          &  0&68422         &  -13&91          \\
        & This work                 & 0&586955(2)     & -15&850(1)      &  0&684209(2)     &  -13&970(1)      \\ \hline

  370 & Ref.~\cite{JiJuan:2008nn} & 0&70531        &  -13&79          &  0&80821         &  -12&00          \\
        & This work                 & 0&705303(2)     & -13&822(1)      &  0&808196(3)     &  -11&986(1)      \\ \hline

  400 & Ref.~\cite{JiJuan:2008nn} & 0&83409        &  -11&75          &  0&9375         &  -9&98           \\
        & This work                 & 0&834083(4)     & -11&765(2)      &  0&937484(4)     &  -9&973(1)       \\ \hline

  450 & Ref.~\cite{JiJuan:2008nn} & 0&95792        &  -9&79           &  1&05294         &  -8&06           \\
        & This work                 & 0&957904(5)     & -9&763(3)       &  1&052917(5)     &  -8&044(2)       \\ \hline

  500 & Ref.~\cite{JiJuan:2008nn} & 1&01384        &  -8&70           &  1&09754         &  -7&09           \\
        & This work                 & 1&013806(6)     & -8&682(4)       &  1&097440(7)     &  -7&064(4)       \\ \hline

  600 & Ref.~\cite{JiJuan:2008nn} & 1&03052        &  -7&77           &  1&09370         &  -6&36           \\
        & This work                 & 1&030489(9)     & -7&714(6)       &  1&093668(9)     &  -6&289(6)       \\ \hline

  700 & Ref.~\cite{JiJuan:2008nn} & 0&99611        &  -7&47           &  1&04437         &  -6&20           \\
        & This work                 & 0&99607(1)      & -7&438(9)       &  1&04437(1)      &  -6&164(9)       \\ \hline

  800 & Ref.~\cite{JiJuan:2008nn} & 0&94567        &  -7&50           &  0&98647         &  -6&61           \\
        & This work                 & 0&94563(1)      & -7&46(1)        &  0&98343(1)      &  -6&30(1)        \\ \hline

  900 & Ref.~\cite{JiJuan:2008nn} & 0&89168        &  -7&71           &  0&92196         &  -6&65           \\
        & This work                 & 0&89164(1)      & -7&62(1)        &  0&92191(1)      &  -6&55(1)        \\ \hline

 1000 & Ref.~\cite{JiJuan:2008nn} & 0&83892        &  -7&94           &  0&86366        &  -6&89           \\
        & This work                 & 0&83887(2)      & -7&86(2)        &  0&86362(2)      &  -6&86(2)        \\ \hline
\end{tabular}
\label{tab-zzz-compa}
\ec
\end{table}

We have also made a comparison for the process $\eewwz$ with the
results of \cite{Wei:2009hq}. In addition to different input
parameters, \cite{Wei:2009hq} uses an unusual scheme/input
parameter for the electromagnetic coupling constant. One can read
from \cite{Wei:2009hq} that their renormalisation condition for
the electric charge is the on-shell definition at $q^2=0$, see
Eq.~(\ref{dze0}), yet the value of $\alpha_{\overline{MS}}(M_Z^2)$
is used as a value for the coupling already at tree level and no
shift of the electric charge counterterm at NLO is made in order
to avoid double counting. This is just like taking a numerical
value of $\alpha_{\overline{MS}}(M_Z^2)$ for $\alpha(0)$ in the
usual on-shell scheme. For the sake of comparison we have used the
same scheme in our calculation and show the comparison with
Table~2 of \cite{Wei:2009hq} in \tab{tab-wwz-compa1}. The
agreement for the Born result is within integration errors as are
the NLO corrections for $M_H = 120\gev$. However for $M_H =
150\gev$ we find agreement only for $\sqrt{s} = 0.5\tev$. Close to
threshold and especially at high energy the results differ by up
to 5 times the integration errors.
\begin{table}[h]
\small
\bc
\caption{\em Born cross section and NLO correction for
  $\eewwz$ using the input parameter scheme of \cite{Wei:2009hq}.}
\vspace*{1ex}
\begin{tabular}{cc@{\qquad}r@{.}l r@{.}l r@{.}l r@{.}l}
\hline
\multicolumn{2}{c}{} & \multicolumn{4}{c}{$M_H=120\gev$} & \multicolumn{4}{c}{$M_H=150\gev$} \\
$\sqrt{s}\,[\tev]$ &  & \multicolumn{2}{c}{$\sigma_{Born}[\fb]$} & \multicolumn{2}{c}{$\Delta\sigma_{NLO}[\fb]$} & \multicolumn{2}{c}{$\sigma_{Born}[\fb]$} & \multicolumn{2}{c}{$\Delta\sigma_{NLO}[\fb]$} \\
    \hline
0.3 & Ref.~\cite{Wei:2009hq} & 3&6216(2) & -0&683(2)& 3&8856(2) & -0&694(2)\\
     & This work & 3&62165(5) & -0&6901(3) & 3&88558(5) & -0&7010(3)\\
    \hline
0.5 & Ref.~\cite{Wei:2009hq} & 44&026(5) & -3&03(6)& 44&303(5) & -2&89(6)\\
     & This work & 44&0235(10) & -3&107(3) & 44&301(1)  & -2&949(3)\\
    \hline
0.8 & Ref.~\cite{Wei:2009hq} & 64&35(1) & -3&48(7)& 64&50(1) & -3&57(9)\\
     & This work & 64&345(4) & -3&466(8) & 64&488(4)  & -3&250(8)\\
    \hline
1.0 & Ref.~\cite{Wei:2009hq} & 65&42(1) & -3&74(9)& 65&51(1) & -3&90(9)\\
     & This work & 65&401(5) & -3&650(9) & 65&499(5)  & -3&440(10)\\
    \hline
\end{tabular}
\label{tab-wwz-compa1}
\ec
\end{table}

\vspace*{0.5cm}

\noi {\bf Note Added}\\
After our paper was made public, \cite{Wei:2009hq} was
updated. The on-shell $\alpha(0)$ scheme is now implemented properly.
Moreover they have improved the numerical stability of the phase-space
integration at high energies and the results for $WWZ$ at $\sqrt{s} = 800 \gev$ and
$1 \tev$ have substantially changed. We now find agreement for these
energies while the small discrepancy near threshold ($\sqrt{s} = 300 \gev$) remains.

\section{Conclusions}

We have performed a calculation of the full next-to-leading order correction to the
processes \eewwzt and \eezzzt in the energy range of the
international linear collider and for Higgs masses below the $WW$
threshold. These processes would be the
successor of \eewwt in that they would measure the quartic
couplings $WWZZ$ and $ZZZZ$ which could retain residual effects of
the physics of electroweak symmetry breaking. With this in mind we
have subtracted the QED corrections and studied the genuine
weak  corrections in the $G_\mu$ scheme.
We find that the weak corrections can be
large. For example, for a centre-of-mass energy of $1\tev$ these
corrections reach $-13\%$ for $WWZ$ and  $-18\%$ for $ZZZ$ and
grow larger for higher energies. At lower energies around
the production threshold the cross sections are small and the
weak corrections are modest.
However, in this energy range the
QED corrections are largest due to the rapid rise of the cross
section. We have also
studied the effects of the genuine weak radiative
corrections on various distributions. While for the $ZZZ$ channel
the effect might be described by an overall rescaling over most
of the range of the kinematic variable under consideration, the
corrections in the $WWZ$ channel show more structure, pointing
once again to the need for radiative corrections when looking for
New Physics
effects.

The calculations involved in our studies are also a
contribution to the field of one-loop corrections for multi-leg
processes and the techniques that one requires to control and
further develop to make these calculations as automatic as
possible. Since these processes involve three-vector boson
production they are much more complex than say single Higgs
production not only from just a counting of the one-loop diagrams
involved but also the fact that the loop integrals are more
challenging, in particular for \eewwztp.
We have shown that using higher precision arithmetic or a combination of
different loop integral libraries can be very efficient to
overcome numerical instabilities.
These techniques are
instrumental for a very precise computation, especially for the
$WWZ$ channel where the foreseen luminosity calls for a better
than per-mil accuracy for the theoretical prediction.
We have shown that our results agree well with
those of a previous calculation of \eezzzt in
\bib{JiJuan:2008nn} but not as well, especially for some Higgs masses
and centre-of-mass energies, for the $WWZ$ channel in \bib{Wei:2009hq}.

\vspace{1cm}
\noi {\bf Acknowledgments} \\
We thank Prof.~Ma Wen-Gan for promptly sending us tables of their results
for a careful comparison. LDN thanks Thomas Hahn for his help with
computers, \fc\ and \lt\ (cache system).
LDN is grateful
to Fukuko Yuasa and Yoshimasa Kurihara for their help with
parallel \bases{}.
This work is part of the
French ANR project, ToolsDMColl BLAN07-2 194882 and is supported
in part by the GDRI-ACPP of the CNRS (France). This work is also
supported in part by the European Community Marie-Curie Research
Training Network under contract MRTN-CT-2006-035505 {\em HEPTOOLS:
Tools and Precision Calculations for Physics Discoveries at
Colliders}.


\end{document}